\documentclass[12pt,english]{article}
\usepackage{mathptmx}
\usepackage[T1]{fontenc}
\usepackage[latin9]{inputenc}
\usepackage{geometry}
\usepackage{color}
\usepackage{babel}
\usepackage{mathtools}
\usepackage{amsmath}
\usepackage{amssymb}
\usepackage{stmaryrd}
\usepackage{graphicx}
\usepackage{setspace}
\usepackage[unicode=true,pdfusetitle,
 bookmarks=true,bookmarksnumbered=false,bookmarksopen=false,
 breaklinks=false,pdfborder={0 0 1},backref=false,colorlinks=true]
 {hyperref}
\hypersetup{
 linkcolor=blue,urlcolor={blue},filecolor={blue},citecolor={blue}}

\makeatletter

\providecommand{\tabularnewline}{\\}

\newcommand{\lyxaddress}[1]{
	\par {\raggedright #1
	\vspace{1.4em}
	\noindent\par}
}

\usepackage{chngcntr}
\usepackage{mathtools}
\counterwithout{figure}{section}
\counterwithout{equation}{section}
\usepackage{times}
\usepackage{amsmath}
\usepackage{floatrow}	
\usepackage{appendix}

\usepackage{times}

\counterwithout{figure}{section}
\counterwithout{equation}{section}
\usepackage{changepage}

\usepackage{babel}

\counterwithout{figure}{section}
\counterwithout{equation}{section}
\usepackage{times}
\usepackage{float}
\usepackage{bm}
\usepackage[caption=false]{subfig}

\usepackage[numbers,sort&compress]{natbib} 

\date{}

\renewcommand{\fnum@figure}{FIG. \thefigure}

\makeatother

\begin{document}
\global\long\def\real{\mathbb{R}}%
 
\global\long\def\RefVol{\Omega_{0}}%
 
\global\long\def\Refx{\mathbf{X}}%
 
\global\long\def\Curx{\mathbf{x}}%
 
\global\long\def\map{\boldsymbol{\chi}}%
 
\global\long\def\defgrad{\mathbf{F}}%
 
\global\long\def\defgradT{\mathbf{F}^{\mathrm{T}}}%
 
\global\long\def\defgradTi{\mathbf{F}^{\mathrm{-T}}}%
 
\global\long\def\d{\mathrm{d}}%
 
\global\long\def\RCG{\mathbf{C}}%
 
\global\long\def\LCG{\mathbf{b}}%
 
\global\long\def\div#1{\nabla\cdot#1}%
 
\global\long\def\curl#1{\nabla\times#1}%
 
\global\long\def\T#1{#1^{\mathrm{T}}}%
 
\global\long\def\CurStress{\boldsymbol{\sigma}}%
 
\global\long\def\Refcurl#1{\nabla_{\mathbf{X}}\times#1}%
 
\global\long\def\s#1{#1^{\star}}%
 
\global\long\def\m#1{#1^{(m)}}%
 
\global\long\def\f#1{#1^{(f)}}%
 
\global\long\def\p#1{#1^{(p)}}%
 
\global\long\def\Refdiv#1{\nabla_{\mathbf{X}}\cdot#1}%
 
\global\long\def\phase#1{#1^{\left(p\right)}}%
 
\global\long\def\jm{j_{m}^{(p)}}%
 
\global\long\def\xinc{\mathbf{\dot{x}}}%
 
\global\long\def\Pinc{\mathbf{\dot{P}}}%
 
\global\long\def\Einc{\mathbf{\dot{E}}}%
 
\global\long\def\Finc{\mathbf{\dot{F}}}%
 
\global\long\def\dinc{\mathbf{\check{d}}}%
 
\global\long\def\einc{\mathbf{\check{e}}}%
 
\global\long\def\Dinc{\mathbf{\dot{D}}}%
 
\global\long\def\sinc{\boldsymbol{\Sigma}}%
 
\global\long\def\fA{\mathbb{\boldsymbol{\mathscr{A}}}}%
 
\global\long\def\fB{\mathbb{\boldsymbol{\mathscr{B}}}}%
 
\global\long\def\fC{\mathbb{\boldsymbol{\mathscr{C}}}}%
 
\global\long\def\plane{\left(x_{1},x_{3}\right)}%
 
\global\long\def\anti{\dot{x}_{2}}%
 
\global\long\def\bnd{\partial\Omega}%
 
\global\long\def\rcg{\mathbf{C}}%
 
\global\long\def\lcg{\mathbf{b}}%
 
\global\long\def\rcgcomp#1{C_{#1}}%
 
\global\long\def\cronck#1{\delta_{#1}}%
 
\global\long\def\lcgcomp#1{b_{#1}}%
 
\global\long\def\deformation{\boldsymbol{\chi}}%
 
\global\long\def\cauchycomp#1{\sigma_{#1}}%
 
\global\long\def\idgt{\dg^{\mathrm{-T}}}%
 
\global\long\def\cauchy{\boldsymbol{\sigma}}%
 
\global\long\def\normal{\mathbf{n}}%
 
\global\long\def\normall{\mathbf{N}}%
 
\global\long\def\traction{\mathbf{t}}%
 
\global\long\def\tractionl{\mathbf{t}_{L}}%
 
\global\long\def\ed{\mathbf{d}}%
 
\global\long\def\edcomp#1{d_{#1}}%
 
\global\long\def\edl{\mathbf{D}}%
 
\global\long\def\edlcomp#1{D_{#1}}%
 
\global\long\def\ef{\mathbf{e}}%
 
\global\long\def\efcomp#1{e_{#1}}%
 
\global\long\def\efl{\mathbf{E}}%
 
\global\long\def\freech{q_{e}}%
 
\global\long\def\surfacech{w_{e}}%
 
\global\long\def\outer#1{#1^{\star}}%
 
\global\long\def\perm{\epsilon_{0}}%
 
\global\long\def\matper{\epsilon}%
 
\global\long\def\jump#1{\llbracket#1\rrbracket}%
 
\global\long\def\identity{\mathbf{I}}%
 
\global\long\def\area{\mathrm{d}a}%
 
\global\long\def\areal{\mathrm{d}A}%
 
\global\long\def\refsys{\mathbf{X}}%
 
\global\long\def\Grad{\nabla_{\refsys}}%
 
\global\long\def\grad{\nabla}%
 
\global\long\def\divg{\nabla\cdot}%
 
\global\long\def\Div{\nabla_{\refsys}}%
 
\global\long\def\derivative#1#2{\frac{\partial#1}{\partial#2}}%
 
\global\long\def\aef{\Psi}%
 
\global\long\def\dltendl{\edl\otimes\edl}%
 
\global\long\def\tr#1{\mathrm{tr}#1}%
 
\global\long\def\ii#1{I_{#1}}%
 
\global\long\def\dh{\hat{D}}%
 
\global\long\def\inc#1{\dot{#1}}%
 
\global\long\def\sys{\mathbf{x}}%
 
\global\long\def\curl{\nabla}%
 
\global\long\def\Curl{\nabla_{\refsys}}%
 
\global\long\def\piolaincpush{\boldsymbol{\Sigma}}%
 
\global\long\def\piolaincpushcomp#1{\Sigma_{#1}}%
 
\global\long\def\elaspush{\boldsymbol{\mathcal{C}}}%
 
\global\long\def\elecpush{\boldsymbol{\mathcal{A}}}%
 
\global\long\def\elaselecpush{\boldsymbol{\mathcal{B}}}%
 
\global\long\def\disgrad{\mathbf{h}}%
 
\global\long\def\disgradcomp#1{h_{#1}}%
 
\global\long\def\trans#1{#1^{\mathrm{\mathsf{T}}}}%
 
\global\long\def\phase#1{#1^{\left(p\right)}}%
 
\global\long\def\elecpushcomp#1{\mathcal{A}_{#1}}%
 
\global\long\def\elaselecpushcomp#1{\mathcal{B}_{#1}}%
 
\global\long\def\elaspushcomp#1{\mathcal{C}_{#1}}%
 
\global\long\def\dnh{\aef_{DH}}%
 
\global\long\def\woo{\varsigma}%
 
\global\long\def\wif{\Lambda}%
 
\global\long\def\structurefun{S}%
 
\global\long\def\dg{\mathbf{F}}%
 
\global\long\def\dgcomp#1{F_{#1}}%
 
\global\long\def\piola{\mathbf{P}}%
 
\global\long\def\refbody{\Omega_{0}}%
 
\global\long\def\refbnd{\partial\refbody}%
 
\global\long\def\bnd{\partial\Omega}%
 
\global\long\def\rcg{\mathbf{C}}%
 
\global\long\def\lcg{\mathbf{b}}%
 
\global\long\def\rcgcomp#1{C_{#1}}%
 
\global\long\def\cronck#1{\delta_{#1}}%
 
\global\long\def\lcgcomp#1{b_{#1}}%
 
\global\long\def\deformation{\boldsymbol{\chi}}%
 
\global\long\def\dgt{\dg^{\mathrm{T}}}%
 
\global\long\def\idgcomp#1{F_{#1}^{-1}}%
 
\global\long\def\velocity{\mathbf{v}}%
 
\global\long\def\accel{\mathbf{a}}%
 
\global\long\def\vg{\mathbf{l}}%
 
\global\long\def\idg{\dg^{-1}}%
 
\global\long\def\cauchycomp#1{\sigma_{#1}}%
 
\global\long\def\idgt{\dg^{\mathrm{-T}}}%
 
\global\long\def\cauchy{\boldsymbol{\sigma}}%
 
\global\long\def\normal{\mathbf{n}}%
 
\global\long\def\normall{\mathbf{N}}%
 
\global\long\def\traction{\mathbf{t}}%
 
\global\long\def\tractionl{\mathbf{t}_{L}}%
 
\global\long\def\ed{\mathbf{d}}%
 
\global\long\def\edcomp#1{d_{#1}}%
 
\global\long\def\edl{\mathbf{D}}%
 
\global\long\def\edlcomp#1{D_{#1}}%
 
\global\long\def\ef{\mathbf{e}}%
 
\global\long\def\efcomp#1{e_{#1}}%
 
\global\long\def\efl{\mathbf{E}}%
 
\global\long\def\freech{q_{e}}%
 
\global\long\def\surfacech{w_{e}}%
 
\global\long\def\outer#1{#1^{\star}}%
 
\global\long\def\perm{\epsilon_{0}}%
 
\global\long\def\matper{\epsilon}%
 
\global\long\def\jump#1{\llbracket#1\rrbracket}%
 
\global\long\def\identity{\mathbf{I}}%
 
\global\long\def\area{\mathrm{d}a}%
 
\global\long\def\areal{\mathrm{d}A}%
 
\global\long\def\refsys{\mathbf{X}}%
 
\global\long\def\Grad{\nabla_{\refsys}}%
 
\global\long\def\grad{\nabla}%
 
\global\long\def\divg{\nabla\cdot}%
 
\global\long\def\Div{\nabla_{\refsys}}%
 
\global\long\def\derivative#1#2{\frac{\partial#1}{\partial#2}}%
 
\global\long\def\aef{\Psi}%
 
\global\long\def\dltendl{\edl\otimes\edl}%
 
\global\long\def\tr#1{\mathrm{tr}\left(#1\right)}%
 
\global\long\def\ii#1{I_{#1}}%
 
\global\long\def\dh{\hat{D}}%
 
\global\long\def\lage{\mathbf{E}}%
 
\global\long\def\inc#1{\dot{#1}}%
 
\global\long\def\sys{\mathbf{x}}%
 
\global\long\def\curl{\nabla}%
 
\global\long\def\Curl{\nabla_{\refsys}}%
 
\global\long\def\piolaincpush{\boldsymbol{\Sigma}}%
 
\global\long\def\piolaincpushcomp#1{\Sigma_{#1}}%
 
\global\long\def\edlincpush{\check{\mathbf{d}}}%
 
\global\long\def\edlincpushcomp#1{\check{d}_{#1}}%
 
\global\long\def\efincpush{\check{\mathbf{e}}}%
 
\global\long\def\efincpushcomp#1{\check{e}_{#1}}%
 
\global\long\def\elaspush{\boldsymbol{\mathcal{C}}}%
 
\global\long\def\elecpush{\boldsymbol{\mathcal{A}}}%
 
\global\long\def\elaselecpush{\boldsymbol{\mathcal{B}}}%
 
\global\long\def\disgrad{\mathbf{h}}%
 
\global\long\def\disgradcomp#1{h_{#1}}%
 
\global\long\def\trans#1{#1^{\mathrm{T}}}%
 
\global\long\def\phase#1{#1^{\left(p\right)}}%
 
\global\long\def\elecpushcomp#1{\mathcal{A}_{#1}}%
 
\global\long\def\elaselecpushcomp#1{\mathcal{B}_{#1}}%
 
\global\long\def\elaspushcomp#1{\mathcal{C}_{#1}}%
 
\global\long\def\dnh{\aef_{DG}}%
 
\global\long\def\dnhc{\mu\lambda^{2}}%
 
\global\long\def\dnhcc{\frac{\mu}{\lambda^{2}}+\frac{1}{\matper}d_{2}^{2}}%
 
\global\long\def\dnhb{\frac{1}{\matper}d_{2}}%
 
\global\long\def\afreq{\omega}%
 
\global\long\def\dispot{\phi}%
 
\global\long\def\edpot{\varphi}%
 
\global\long\def\kh{\hat{k}}%
 
\global\long\def\afreqh{\hat{\afreq}}%
 
\global\long\def\phasespeed{c}%
 
\global\long\def\bulkspeed{c_{B}}%
 
\global\long\def\speedh{\hat{c}}%
 
\global\long\def\dhth{\dh_{th}}%
 
\global\long\def\bulkspeedl{\bulkspeed_{\lambda}}%
 
\global\long\def\khth{\hat{k}_{th}}%
 
\global\long\def\p#1{#1^{\left(p\right)}}%
 
\global\long\def\maxinccomp#1{\inc{\outer{\sigma}}_{#1}}%
 
\global\long\def\maxcomp#1{\outer{\sigma}_{#1}}%
 
\global\long\def\relper{\matper_{r}}%
 
\global\long\def\sdh{\hat{d}}%
 
\global\long\def\iee{\varphi}%
 
\global\long\def\effectivemu{\tilde{\mu}}%
 
\global\long\def\fb#1{#1^{\left(a\right)}}%
 
\global\long\def\mt#1{#1^{\left(b\right)}}%
 
\global\long\def\phs#1{#1^{\left(p\right)}}%
 
\global\long\def\thc{h}%
 
\global\long\def\state{\mathbf{s}}%
 
\global\long\def\harmonicper{\breve{\matper}}%
 
\global\long\def\kb{k_{B}}%
 
\global\long\def\cb{\bar{c}}%
 
\global\long\def\mb{\bar{\mu}}%
 
\global\long\def\rb{\bar{\rho}}%
 
\global\long\def\wavenumber{k}%
 
\global\long\def\nh{\mathbf{n}}%
 
\global\long\def\mh{\mathbf{m}}%
 
\global\long\def\deflect{\inc x_{2}}%
 
\global\long\def\sdd#1{#1_{2,11}}%
 
\global\long\def\sdddd#1{#1_{2,1111}}%
 
\global\long\def\sd#1{#1_{2,1}}%
 
\global\long\def\sddd#1{#1_{2,111}}%
 
\global\long\def\xdddd#1{#1_{,\xi\xi\xi\xi}}%
 
\global\long\def\xdd#1{#1_{,\xi\xi}}%
 
\global\long\def\xd#1{#1_{,\xi}}%
 
\global\long\def\xddd#1{#1_{,\xi\xi\xi}}%
 
\global\long\def\jm{J_{m}}%
 
\global\long\def\dv{\Delta V}%
 
\global\long\def\ih{\mathbf{i}_{1}}%
 
\global\long\def\kh{\mathbf{i}_{3}}%
 
\global\long\def\jh{\mathbf{i}_{2}}%
 
\global\long\def\etil{E}%
 
\global\long\def\genT{\mathsf{Q}}%
 
\global\long\def\transfer{\mathsf{T}}%
 
\global\long\def\statevec{\mathbf{s}}%
 
\global\long\def\coefvec{\mathbf{c}}%
 
\global\long\def\pressure{p_{0}}%
 
\global\long\def\ncell#1{#1_{\left(n\right)}}%
 
\global\long\def\ydisp{\inc x_{2}}%
 
\global\long\def\ycord{x_{2}}%
 
\global\long\def\pn#1{\ncell{#1}^{\left(p\right)}}%
 
\global\long\def\pnm#1{#1_{\left(n\right)m}^{\left(p\right)}}%
 
\global\long\def\eigen{\boldsymbol{\eta}}%
 
\global\long\def\xcomp{x_{1}}%
 
\global\long\def\totalT{\mathsf{T_{\mathrm{tot}}}}%
 
\global\long\def\rads{\frac{\mathrm{rad}}{\mathrm{s}}}%
 
\global\long\def\lf{\gamma}%
 
\global\long\def\tf{T_{m}}%
 
\global\long\def\eigenim{\beta}%
 
\global\long\def\bS{\mathsf{S}}%

\global\long\def\Blochwn{{k}_{\mathit{B}}}%
 
\global\long\def\Blochwnx{k_{B1}}%
 
\global\long\def\Blochwny{k_{B2}}%
 
\global\long\def\waven{k}%

\global\long\def\Beff{\tilde{B}}%
 
\global\long\def\rhoaux{\check{\rho}}%
 
\global\long\def\nuaux{\check{\nu}}%
 
\global\long\def\Daux{\check{D}}%

\global\long\def\geff{\tilde{g}}%
 
\global\long\def\avevarsigma{\overline{\varsigma}}%

\global\long\def\Deff{\tilde{D}}%
 
\global\long\def\Dxeff{\tilde{D}_{1}}%
 
\global\long\def\Dyeff{\tilde{D}_{2}}%
 
\global\long\def\Dnueff{\tilde{D}_{\nu}}%
 
\global\long\def\Dxyeff{\tilde{D}_{12}}%
 
\global\long\def\nueff{\tilde{\nu}}%
 
\global\long\def\rhoeff{\tilde{\rho}}%

\global\long\def\aveR{\overline{R}}%
 
\global\long\def\aveMx{\overline{M}_{1}}%
 
\global\long\def\aveMy{\overline{M}_{2}}%
 
\global\long\def\aveMxy{\overline{M}_{12}}%
 
\global\long\def\aveM{\overline{M}}%
 
\global\long\def\avew{\overline{w}}%
 
\global\long\def\avep{\overline{p}}%

\global\long\def\vectora{\boldsymbol{\psi}}%
 
\global\long\def\vectorb{\boldsymbol{\varphi}}%
 
\global\long\def\vectorp{\mathsf{P}}%

\global\long\def\avevectora{\overline{\boldsymbol{\psi}} }%
 
\global\long\def\avevectorao{\left\langle \boldsymbol{\psi}_{0}\right\rangle }%
 
\global\long\def\avevectorb{\overline{\boldsymbol{\varphi}} }%
 
\global\long\def\avevectorp{\overline{\mathsf{P}} }%

\global\long\def\avevectorax{\left\langle \boldsymbol{\psi}{_{,x}}\right\rangle }%
 
\global\long\def\avevectorat{\left\langle \boldsymbol{\psi}{_{,t}}\right\rangle }%
 
\global\long\def\avevectorbx{\left\langle \boldsymbol{\varphi}{_{,x}}\right\rangle }%

\global\long\def\DDeff{\tilde{\tilde{D}}}%
 
\global\long\def\DDxeff{\tilde{\tilde{D}}_{1}}%
 
\global\long\def\DDyeff{\tilde{\tilde{D}}_{2}}%
 
\global\long\def\DDnueff{\tilde{\tilde{D}}_{\nu}}%
 
\global\long\def\nuueff{\tilde{\tilde{\nu}}}%
 
\global\long\def\DDkeff{\tilde{\tilde{D}}_{k}}%
 
\global\long\def\DDxyeff{\tilde{\tilde{D}}_{12}}%
 
\global\long\def\rhooeff{\tilde{\tilde{\rho}}}%

\global\long\def\angRR{\left\langle \left\langle R_{\mathrm{p}}\right\rangle \right\rangle }%
 
\global\long\def\angMMx{\left\langle M_{1\mathrm{p}}\right\rangle }%
 
\global\long\def\angMMy{\left\langle M_{2\mathrm{p}}\right\rangle }%
 
\global\long\def\angMMxy{\left\langle M_{12\mathrm{p}}\right\rangle }%
 
\global\long\def\angww{\left\langle w_{\mathrm{p}}\right\rangle }%
 
\global\long\def\angpp{ \left\langle p_{\mathrm{p}}\right\rangle }%

\global\long\def\uF{u_{\mathrm{F}}}%
 
\global\long\def\wF{w_{\mathrm{F}}}%
 
\global\long\def\rhoF{\rho_{\mathrm{F}}}%
 
\global\long\def\BF{B_{\mathrm{F}}}%
 
\global\long\def\dl{l^{(i)}}%

\global\long\def\bF{\state_{\mathrm{d}}}%
 
\global\long\def\bA{\state_{\mathrm{f}}}%
 
\global\long\def\ba{\coefvec}%
 
\global\long\def\bcero{\mathsf{0}}%
 
\global\long\def\bH{\mathsf{H}}%
 
\global\long\def\ltotal{l_{\mathrm{tot}}}%
 
\global\long\def\bI{\mathsf{I}}%
 
\global\long\def\qpd{\genT_{\mathrm{d}}^{+}}%
 
\global\long\def\qmd{\genT_{\mathrm{d}}^{-}}%
 
\global\long\def\qpf{\genT_{\mathrm{f}}^{+}}%
 
\global\long\def\qmf{\genT_{\mathrm{f}}^{-}}%
 
\global\long\def\curcon{\Omega}%
 
\global\long\def\ld{\mathbf{D}}%
 
\global\long\def\qpdz{\genT_{\mathrm{d}\left(0\right)}^{+}}%
 
\global\long\def\qmdz{\genT_{\mathrm{d}\left(0\right)}^{-}}%
 
\global\long\def\qpfz{\genT_{\mathrm{f}\left(0\right)}^{+}}%
 
\global\long\def\qmfz{\genT_{\mathrm{f}\left(0\right)}^{-}}%
 
\global\long\def\qpdl{\genT_{\mathrm{d}\left(M\right)}^{+}}%
 
\global\long\def\qmdl{\genT_{\mathrm{d}\left(M\right)}^{-}}%
 
\global\long\def\qpfl{\genT_{\mathrm{f}\left(M\right)}^{+}}%
 
\global\long\def\qmfl{\genT_{\mathrm{f}\left(M\right)}^{-}}%
 
\global\long\def\e{\mathop{{\rm \mbox{{\Large e}}}}\nolimits}%
 
\global\long\def\Tr{\textrm{Tr}}%
 
\global\long\def\Det{\textrm{Det}}%
 
\global\long\def\sgn{\textrm{sgn}}%
 
\global\long\def\pr{^{\prime}}%
 
\global\long\def\bn#1{\mbox{\boldmath\ensuremath{#1}}}%
 
\global\long\def\bB{\bn B}%
 
\global\long\def\bP{\bn P}%
 
\global\long\def\bY{\bn Y}%
 
\global\long\def\bV{\bn V}%
 
\global\long\def\bW{\bn W}%
 
\global\long\def\bG{\bn G}%

\global\long\def\bM{\bn M}%
 
\global\long\def\bm{\bn m}%
 
\global\long\def\bE{\bn E}%
 
\global\long\def\bK{\bn K}%
 
\global\long\def\bL{\bn L}%
 
\global\long\def\bC{\bn C}%
 
\global\long\def\bT{\bn T}%
 
\global\long\def\bg{\bn g}%
 
\global\long\def\bN{\bn N}%
 
\global\long\def\bX{\bn X}%
 
\global\long\def\bR{\bn R}%
 
\global\long\def\bHs{\bn{Hs}}%
 
\global\long\def\bDelta{\bn{\Delta}}%
 
\global\long\def\bsigma{\bn{\sigma}}%
 
\global\long\def\bpsi{\bn{\psi}}%
 
\global\long\def\bQ{\bn Q}%
 
\global\long\def\bZ{\bn Z}%
 
\global\long\def\bU{\bn U}%
 
\global\long\def\bz{\bn z}%
 
\global\long\def\bu{\bn u}%
 
\global\long\def\bk{\bn k}%
 
\global\long\def\bUpsilon{\bn{\Upsilon}}%
 
\global\long\def\bEta{\bn{\eta}}%
 
\global\long\def\bmu{\bn{\mu}}%
 
\global\long\def\brho{\bn{\rho}}%
 
\global\long\def\bpartial{\bn{\partial}}%

\global\long\def\Elecf{\mathbf{E}}%
 
\global\long\def\Edisplacement{\mathbf{D}}%
 
\global\long\def\Strain{\grad\disp}%
 
\global\long\def\Stress{\boldsymbol{\sigma}}%
 
\global\long\def\Piezoelectricmodule{\mathbf{B}}%
 
\global\long\def\Permittivity{\mathbf{A}}%

\global\long\def\alfacoupled{\check{\alpha}}%
 
\global\long\def\gammacoupled{\check{\gamma}}%
 
\global\long\def\bettacoupled{\check{\beta}}%
 
\global\long\def\rhocoupled{\check{\rho}}%
 
\global\long\def\Scoupled{\check{S}}%
 
\global\long\def\Lcoupled{\check{L}}%
 
\global\long\def\Qcoupled{\check{Q}}%

\global\long\def\elas{\mathbf{C}}%
 
\global\long\def\momentum{\mathbf{p}}%
 
\global\long\def\force{\mathbf{f}}%
 
\global\long\def\disp{\mathbf{u}}%
 
\global\long\def\potential{\phi}%
 
\global\long\def\state{\mathsf{S}}%
 
\global\long\def\position{\mathbf{x}}%
 
\global\long\def\cmat{\mathsf{C}}%
 
\global\long\def\rmat{\mathsf{R}}%
 
\global\long\def\mmat{\mathsf{m}}%
 
\global\long\def\pmat{\mathsf{p}}%
 
\global\long\def\gradt{\mathsf{B}}%
 
\global\long\def\genpotential{\mathsf{w}}%
 
\global\long\def\vol{\Omega}%
 
\global\long\def\rve{\Omega^{\mathrm{rve}}}%
 
\global\long\def\effective#1{\tilde{#1}}%
\global\long\def\ensemble#1{\left\langle #1\right\rangle }%
\global\long\def\average#1{\overline{#1}}%
\global\long\def\rg{\mathbf{W}}%
\global\long\def\po#1{#1^{\left(1\right)}}%
 
\global\long\def\pt#1{#1^{\left(2\right)}}%
 
\global\long\def\ptr#1{#1^{\left(3\right)}}%
 
\global\long\def\length{l}%
\global\long\def\dt#1{\dot{#1}}%
\global\long\def\willis{\mathbf{S}}%
\global\long\def\area{a}%
\global\long\def\fmat{\mathsf{f}}%
\global\long\def\lmat{\mathsf{L}}%
\global\long\def\hmat{\mathsf{h}}%
\global\long\def\bmat{\mathsf{b}}%
 
\global\long\def\green{\mathsf{G}}%
\global\long\def\adjoint#1{#1^{\dagger}}%
\global\long\def\cc#1{#1^{*}}%
\global\long\def\intvol#1{\int_{\vol}#1\mathrm{d}\vol}%
\global\long\def\boundaryw{\bnd_{\mathsf{w}}}%
\global\long\def\intsurfacepot#1{\int_{\boundaryw}#1\mathrm{d}\area}%
\global\long\def\boundaryt{\bnd_{\mathsf{t}}}%
\global\long\def\intsurfaceh#1{\int_{\boundaryt}#1\mathrm{d}\area}%
\global\long\def\normalop{\mathsf{n}}%
\global\long\def\smat{\mathsf{s}}%
\global\long\def\fmat{\mathsf{f}}%
\global\long\def\dmat{\mathsf{D}}%
\global\long\def\zm#1#2{\mathsf{0}_{#1\times#2}}%
\global\long\def\nmat{\mathsf{N}}%
\global\long\def\tp#1{#1^{\mathsf{T}}}%
\global\long\def\charge{q}%
\global\long\def\scharge{\omega_{\mathrm{c}}}%
\global\long\def\dirac{\delta}%
\global\long\def\period{l}%
\global\long\def\scalarf{f}%
\global\long\def\scalarD{D}%
\global\long\def\randomvar{p}%
\global\long\def\cell{\vol_{\mathrm{p}}}%
\global\long\def\realization{y}%
\global\long\def\realizationf{\zeta_{\realization}}%
\global\long\def\periodicf{\zeta_{\mathrm{p}}}%
\global\long\def\realizations#1{#1_{y}}%
\global\long\def\ensembleint#1{\int_{\cell}#1\mathrm{d}\realization}%
\global\long\def\periodic#1{#1_{\mathrm{p}}}%
\global\long\def\pl#1{#1^{+}}%
\global\long\def\mn#1{#1^{-}}%
\global\long\def\scalarG{G}%
\global\long\def\zerosub#1{#1_{0}}%
\global\long\def\parameter{y}%
\global\long\def\samplespace{\Lambda}%
\global\long\def\scalarelas{C}%
\global\long\def\scalarpiezo{B}%
\global\long\def\scalarA{A}%
\global\long\def\scalarS{S}%
\global\long\def\scalarW{W}%
\global\long\def\effectivemass{\tilde{\boldsymbol{\rho}}}%

\global\long\def\constsub#1{#1_{c}}%
\global\long\def\wronskian{V}%
\global\long\def\eigens{\eta}%
\global\long\def\intsurfacetotal#1{\int_{\bnd}#1\mathrm{d}\area}%

\global\long\def\real{\mathbb{R}}%
\global\long\def\RefVol{\Omega_{0}}%
\global\long\def\Refx{\mathbf{X}}%
\global\long\def\Curx{\mathbf{x}}%
\global\long\def\map{\boldsymbol{\chi}}%
\global\long\def\defgrad{\mathbf{F}}%
\global\long\def\defgradT{\mathbf{F}^{\mathrm{T}}}%
\global\long\def\defgradTi{\mathbf{F}^{\mathrm{-T}}}%
\global\long\def\d{\mathrm{d}}%
\global\long\def\RCG{\mathbf{C}}%
\global\long\def\LCG{\mathbf{b}}%
\global\long\def\div#1{\nabla\cdot#1}%
\global\long\def\curl#1{\nabla\times#1}%
\global\long\def\T#1{#1^{\mathrm{T}}}%
\global\long\def\CurStress{\boldsymbol{\sigma}}%
\global\long\def\Refcurl#1{\nabla_{\mathbf{X}}\times#1}%
\global\long\def\s#1{#1^{\star}}%
\global\long\def\m#1{#1^{(m)}}%
\global\long\def\f#1{#1^{(f)}}%
\global\long\def\p#1{#1^{(p)}}%
\global\long\def\Refdiv#1{\nabla_{\mathbf{X}}\cdot#1}%
\global\long\def\phase#1{#1^{\left(p\right)}}%
\global\long\def\jm{j_{m}^{(p)}}%
\global\long\def\xinc{\mathbf{\dot{x}}}%
\global\long\def\Pinc{\mathbf{\dot{P}}}%
\global\long\def\Einc{\mathbf{\dot{E}}}%
\global\long\def\Finc{\mathbf{\dot{F}}}%
\global\long\def\dinc{\mathbf{\check{d}}}%
\global\long\def\einc{\mathbf{\check{e}}}%
\global\long\def\Dinc{\mathbf{\dot{D}}}%
\global\long\def\sinc{\boldsymbol{\Sigma}}%
\global\long\def\fA{\mathbb{\boldsymbol{\mathscr{A}}}}%
\global\long\def\fB{\mathbb{\boldsymbol{\mathscr{B}}}}%
\global\long\def\fC{\mathbb{\boldsymbol{\mathscr{C}}}}%
\global\long\def\plane{\left(x_{1},x_{3}\right)}%
\global\long\def\anti{\dot{x}_{2}}%
\global\long\def\bnd{\partial\Omega}%
\global\long\def\rcg{\mathbf{C}}%
\global\long\def\lcg{\mathbf{b}}%
\global\long\def\rcgcomp#1{C_{#1}}%
\global\long\def\cronck#1{\delta_{#1}}%
\global\long\def\lcgcomp#1{b_{#1}}%
\global\long\def\deformation{\boldsymbol{\chi}}%
\global\long\def\cauchycomp#1{\sigma_{#1}}%
\global\long\def\idgt{\dg^{\mathrm{-T}}}%
\global\long\def\cauchy{\boldsymbol{\sigma}}%
\global\long\def\normal{\mathbf{n}}%
\global\long\def\normall{\mathbf{N}}%
\global\long\def\traction{\mathbf{t}}%
\global\long\def\tractionl{\mathbf{t}_{L}}%
\global\long\def\ed{\mathbf{d}}%
\global\long\def\edcomp#1{d_{#1}}%
\global\long\def\edl{\mathbf{D}}%
\global\long\def\edlcomp#1{D_{#1}}%
\global\long\def\ef{\mathbf{e}}%
\global\long\def\efcomp#1{e_{#1}}%
\global\long\def\efl{\mathbf{E}}%
\global\long\def\freech{q_{e}}%
\global\long\def\surfacech{w_{e}}%
\global\long\def\outer#1{#1^{\star}}%
\global\long\def\perm{\epsilon_{0}}%
\global\long\def\matper{\epsilon}%
\global\long\def\jump#1{\llbracket#1\rrbracket}%
\global\long\def\identity{\mathbf{I}}%
\global\long\def\area{\mathrm{d}a}%
\global\long\def\areal{\mathrm{d}A}%
\global\long\def\refsys{\mathbf{X}}%
\global\long\def\Grad{\nabla_{\refsys}}%
\global\long\def\grad{\nabla}%
\global\long\def\divg{\nabla\cdot}%
\global\long\def\Div{\nabla_{\refsys}}%
\global\long\def\derivative#1#2{\frac{\partial#1}{\partial#2}}%
\global\long\def\aef{\Psi}%
\global\long\def\dltendl{\edl\otimes\edl}%
\global\long\def\tr#1{\mathrm{tr}#1}%
\global\long\def\ii#1{I_{#1}}%
\global\long\def\dh{\hat{D}}%
\global\long\def\inc#1{\dot{#1}}%
\global\long\def\sys{\mathbf{x}}%
\global\long\def\curl{\nabla}%
\global\long\def\Curl{\nabla_{\refsys}}%
\global\long\def\piolaincpush{\boldsymbol{\Sigma}}%
\global\long\def\piolaincpushcomp#1{\Sigma_{#1}}%
\global\long\def\elaspush{\boldsymbol{\mathcal{C}}}%
\global\long\def\elecpush{\boldsymbol{\mathcal{A}}}%
\global\long\def\elaselecpush{\boldsymbol{\mathcal{B}}}%
\global\long\def\disgrad{\mathbf{h}}%
\global\long\def\disgradcomp#1{h_{#1}}%
\global\long\def\trans#1{#1^{\mathrm{\mathsf{T}}}}%
\global\long\def\phase#1{#1^{\left(p\right)}}%
\global\long\def\elecpushcomp#1{\mathcal{A}_{#1}}%
\global\long\def\elaselecpushcomp#1{\mathcal{B}_{#1}}%
\global\long\def\elaspushcomp#1{\mathcal{C}_{#1}}%
\global\long\def\dnh{\aef_{DH}}%
\global\long\def\woo{\varsigma}%
\global\long\def\wif{\Lambda}%
\global\long\def\structurefun{S}%
\global\long\def\dg{\mathbf{F}}%
\global\long\def\dgcomp#1{F_{#1}}%
\global\long\def\piola{\mathbf{P}}%
\global\long\def\refbody{\Omega_{0}}%
\global\long\def\refbnd{\partial\refbody}%
\global\long\def\bnd{\partial\Omega}%
\global\long\def\rcg{\mathbf{C}}%
\global\long\def\lcg{\mathbf{b}}%
\global\long\def\rcgcomp#1{C_{#1}}%
\global\long\def\cronck#1{\delta_{#1}}%
\global\long\def\lcgcomp#1{b_{#1}}%
\global\long\def\deformation{\boldsymbol{\chi}}%
\global\long\def\dgt{\dg^{\mathrm{T}}}%
\global\long\def\idgcomp#1{F_{#1}^{-1}}%
\global\long\def\velocity{\mathbf{v}}%
\global\long\def\accel{\mathbf{a}}%
\global\long\def\vg{\mathbf{l}}%
\global\long\def\idg{\dg^{-1}}%
\global\long\def\cauchycomp#1{\sigma_{#1}}%
\global\long\def\idgt{\dg^{\mathrm{-T}}}%
\global\long\def\cauchy{\boldsymbol{\sigma}}%
\global\long\def\normal{\mathbf{n}}%
\global\long\def\normall{\mathbf{N}}%
\global\long\def\traction{\mathbf{t}}%
\global\long\def\tractionl{\mathbf{t}_{L}}%
\global\long\def\ed{\mathbf{d}}%
\global\long\def\edcomp#1{d_{#1}}%
\global\long\def\edl{\mathbf{D}}%
\global\long\def\edlcomp#1{D_{#1}}%
\global\long\def\ef{\mathbf{e}}%
\global\long\def\efcomp#1{e_{#1}}%
\global\long\def\efl{\mathbf{E}}%
\global\long\def\freech{q_{e}}%
\global\long\def\surfacech{w_{e}}%
\global\long\def\outer#1{#1^{\star}}%
\global\long\def\perm{\epsilon_{0}}%
\global\long\def\matper{\epsilon}%
\global\long\def\jump#1{\llbracket#1\rrbracket}%
\global\long\def\identity{\mathbf{I}}%
\global\long\def\area{\mathrm{d}a}%
\global\long\def\areal{\mathrm{d}A}%
\global\long\def\refsys{\mathbf{X}}%
\global\long\def\Grad{\nabla_{\refsys}}%
\global\long\def\grad{\nabla}%
\global\long\def\divg{\nabla\cdot}%
\global\long\def\Div{\nabla_{\refsys}}%
\global\long\def\derivative#1#2{\frac{\partial#1}{\partial#2}}%
\global\long\def\aef{\Psi}%
\global\long\def\dltendl{\edl\otimes\edl}%
\global\long\def\tr#1{\mathrm{tr}\left(#1\right)}%
\global\long\def\ii#1{I_{#1}}%
\global\long\def\dh{\hat{D}}%
\global\long\def\lage{\mathbf{E}}%
\global\long\def\inc#1{\dot{#1}}%
\global\long\def\sys{\mathbf{x}}%
\global\long\def\curl{\nabla}%
\global\long\def\Curl{\nabla_{\refsys}}%
\global\long\def\piolaincpush{\boldsymbol{\Sigma}}%
\global\long\def\piolaincpushcomp#1{\Sigma_{#1}}%
\global\long\def\edlincpush{\check{\mathbf{d}}}%
\global\long\def\edlincpushcomp#1{\check{d}_{#1}}%
\global\long\def\efincpush{\check{\mathbf{e}}}%
\global\long\def\efincpushcomp#1{\check{e}_{#1}}%
\global\long\def\elaspush{\boldsymbol{\mathcal{C}}}%
\global\long\def\elecpush{\boldsymbol{\mathcal{A}}}%
\global\long\def\elaselecpush{\boldsymbol{\mathcal{B}}}%
\global\long\def\disgrad{\mathbf{h}}%
\global\long\def\disgradcomp#1{h_{#1}}%
\global\long\def\trans#1{#1^{\mathrm{T}}}%
\global\long\def\phase#1{#1^{\left(p\right)}}%
\global\long\def\elecpushcomp#1{\mathcal{A}_{#1}}%
\global\long\def\elaselecpushcomp#1{\mathcal{B}_{#1}}%
\global\long\def\elaspushcomp#1{\mathcal{C}_{#1}}%
\global\long\def\dnh{\aef_{DG}}%
\global\long\def\dnhc{\mu\lambda^{2}}%
\global\long\def\dnhcc{\frac{\mu}{\lambda^{2}}+\frac{1}{\matper}d_{2}^{2}}%
\global\long\def\dnhb{\frac{1}{\matper}d_{2}}%
\global\long\def\afreq{\omega}%
\global\long\def\dispot{\phi}%
\global\long\def\edpot{\varphi}%
\global\long\def\kh{\hat{k}}%
\global\long\def\afreqh{\hat{\afreq}}%
\global\long\def\phasespeed{c}%
\global\long\def\bulkspeed{c_{B}}%
\global\long\def\speedh{\hat{c}}%
\global\long\def\dhth{\dh_{th}}%
\global\long\def\bulkspeedl{\bulkspeed_{\lambda}}%
\global\long\def\khth{\hat{k}_{th}}%
\global\long\def\p#1{#1^{\left(p\right)}}%
\global\long\def\maxinccomp#1{\inc{\outer{\sigma}}_{#1}}%
\global\long\def\maxcomp#1{\outer{\sigma}_{#1}}%
\global\long\def\relper{\matper_{r}}%
\global\long\def\sdh{\hat{d}}%
\global\long\def\iee{\varphi}%
\global\long\def\effectivemu{\tilde{\mu}}%
\global\long\def\fb#1{#1^{\left(a\right)}}%
\global\long\def\mt#1{#1^{\left(b\right)}}%
\global\long\def\phs#1{#1^{\left(p\right)}}%
\global\long\def\thc{h}%
\global\long\def\state{\mathbf{s}}%
\global\long\def\harmonicper{\breve{\matper}}%
\global\long\def\kb{k_{B}}%
\global\long\def\cb{\bar{c}}%
\global\long\def\mb{\bar{\mu}}%
\global\long\def\rb{\bar{\rho}}%
\global\long\def\wavenumber{k}%
\global\long\def\nh{\mathbf{n}}%
\global\long\def\mh{\mathbf{m}}%
\global\long\def\deflect{\inc x_{2}}%
\global\long\def\sdd#1{#1_{2,11}}%
\global\long\def\sdddd#1{#1_{2,1111}}%
\global\long\def\sd#1{#1_{2,1}}%
\global\long\def\sddd#1{#1_{2,111}}%
\global\long\def\xdddd#1{#1_{,\xi\xi\xi\xi}}%
\global\long\def\xdd#1{#1_{,\xi\xi}}%
\global\long\def\xd#1{#1_{,\xi}}%
\global\long\def\xddd#1{#1_{,\xi\xi\xi}}%
\global\long\def\jm{J_{m}}%
\global\long\def\dv{\Delta V}%
\global\long\def\ih{\mathbf{i}_{1}}%
\global\long\def\kh{\mathbf{i}_{3}}%
\global\long\def\jh{\mathbf{i}_{2}}%
\global\long\def\etil{E}%
\global\long\def\genT{\mathsf{Q}}%
\global\long\def\transfer{\mathsf{T}}%
\global\long\def\statevec{\mathbf{s}}%
\global\long\def\coefvec{\mathbf{c}}%
\global\long\def\pressure{p_{0}}%
\global\long\def\ncell#1{#1_{\left(n\right)}}%
\global\long\def\ydisp{\inc x_{2}}%
\global\long\def\ycord{x_{2}}%
\global\long\def\pn#1{\ncell{#1}^{\left(p\right)}}%
\global\long\def\pnm#1{#1_{\left(n\right)m}^{\left(p\right)}}%
\global\long\def\eigen{\boldsymbol{\eta}}%
\global\long\def\xcomp{x_{1}}%
\global\long\def\totalT{\mathsf{T_{\mathrm{tot}}}}%
\global\long\def\rads{\frac{\mathrm{rad}}{\mathrm{s}}}%
\global\long\def\lf{\gamma}%
\global\long\def\tf{T_{m}}%
\global\long\def\eigenim{\beta}%
\global\long\def\bS{\mathsf{S}}%
\global\long\def\Blochwn{{k}_{\mathit{B}}}%
\global\long\def\Blochwnx{k_{B1}}%
\global\long\def\Blochwny{k_{B2}}%
\global\long\def\waven{k}%
\global\long\def\Beff{\tilde{B}}%
\global\long\def\rhoaux{\check{\rho}}%
\global\long\def\nuaux{\check{\nu}}%
\global\long\def\Daux{\check{D}}%
\global\long\def\geff{\tilde{g}}%
\global\long\def\avevarsigma{\overline{\varsigma}}%
\global\long\def\Deff{\tilde{D}}%
\global\long\def\Dxeff{\tilde{D}_{1}}%
\global\long\def\Dyeff{\tilde{D}_{2}}%
\global\long\def\Dnueff{\tilde{D}_{\nu}}%
\global\long\def\Dxyeff{\tilde{D}_{12}}%
\global\long\def\nueff{\tilde{\nu}}%
\global\long\def\rhoeff{\tilde{\rho}}%
\global\long\def\aveR{\overline{R}}%
\global\long\def\aveMx{\overline{M}_{1}}%
\global\long\def\aveMy{\overline{M}_{2}}%
\global\long\def\aveMxy{\overline{M}_{12}}%
\global\long\def\aveM{\overline{M}}%
\global\long\def\avew{\overline{w}}%
\global\long\def\avep{\overline{p}}%
\global\long\def\vectora{\boldsymbol{\psi}}%
\global\long\def\vectorb{\boldsymbol{\varphi}}%
\global\long\def\vectorp{\mathsf{P}}%
\global\long\def\avevectora{\overline{\boldsymbol{\psi}} }%
\global\long\def\avevectorao{\left\langle \boldsymbol{\psi}_{0}\right\rangle }%
\global\long\def\avevectorb{\overline{\boldsymbol{\varphi}} }%
\global\long\def\avevectorp{\overline{\mathsf{P}} }%
\global\long\def\avevectorax{\left\langle \boldsymbol{\psi}{_{,x}}\right\rangle }%
\global\long\def\avevectorat{\left\langle \boldsymbol{\psi}{_{,t}}\right\rangle }%
\global\long\def\avevectorbx{\left\langle \boldsymbol{\varphi}{_{,x}}\right\rangle }%
\global\long\def\DDeff{\tilde{\tilde{D}}}%
\global\long\def\DDxeff{\tilde{\tilde{D}}_{1}}%
\global\long\def\DDyeff{\tilde{\tilde{D}}_{2}}%
\global\long\def\DDnueff{\tilde{\tilde{D}}_{\nu}}%
\global\long\def\nuueff{\tilde{\tilde{\nu}}}%
\global\long\def\DDkeff{\tilde{\tilde{D}}_{k}}%
\global\long\def\DDxyeff{\tilde{\tilde{D}}_{12}}%
\global\long\def\rhooeff{\tilde{\tilde{\rho}}}%
\global\long\def\angRR{\left\langle \left\langle R_{\mathrm{p}}\right\rangle \right\rangle }%
\global\long\def\angMMx{\left\langle M_{1\mathrm{p}}\right\rangle }%
\global\long\def\angMMy{\left\langle M_{2\mathrm{p}}\right\rangle }%
\global\long\def\angMMxy{\left\langle M_{12\mathrm{p}}\right\rangle }%
\global\long\def\angww{\left\langle w_{\mathrm{p}}\right\rangle }%
\global\long\def\angpp{ \left\langle p_{\mathrm{p}}\right\rangle }%
\global\long\def\uF{u_{\mathrm{F}}}%
\global\long\def\wF{w_{\mathrm{F}}}%
\global\long\def\rhoF{\rho_{\mathrm{F}}}%
\global\long\def\BF{B_{\mathrm{F}}}%
\global\long\def\dl{l^{(i)}}%
\global\long\def\bF{\state_{\mathrm{d}}}%
\global\long\def\bA{\state_{\mathrm{f}}}%
\global\long\def\ba{\coefvec}%
\global\long\def\bcero{\mathsf{0}}%
\global\long\def\bH{\mathsf{H}}%
\global\long\def\ltotal{l_{\mathrm{tot}}}%
\global\long\def\bI{\mathsf{I}}%
\global\long\def\qpd{\genT_{\mathrm{d}}^{+}}%
\global\long\def\qmd{\genT_{\mathrm{d}}^{-}}%
\global\long\def\qpf{\genT_{\mathrm{f}}^{+}}%
\global\long\def\qmf{\genT_{\mathrm{f}}^{-}}%
\global\long\def\curcon{\Omega}%
\global\long\def\ld{\mathbf{D}}%
\global\long\def\qpdz{\genT_{\mathrm{d}\left(0\right)}^{+}}%
\global\long\def\qmdz{\genT_{\mathrm{d}\left(0\right)}^{-}}%
\global\long\def\qpfz{\genT_{\mathrm{f}\left(0\right)}^{+}}%
\global\long\def\qmfz{\genT_{\mathrm{f}\left(0\right)}^{-}}%
\global\long\def\qpdl{\genT_{\mathrm{d}\left(M\right)}^{+}}%
\global\long\def\qmdl{\genT_{\mathrm{d}\left(M\right)}^{-}}%
\global\long\def\qpfl{\genT_{\mathrm{f}\left(M\right)}^{+}}%
\global\long\def\qmfl{\genT_{\mathrm{f}\left(M\right)}^{-}}%
\global\long\def\e{\mathop{{\rm \mbox{{\Large e}}}}\nolimits}%
\global\long\def\Tr{\textrm{Tr}}%
\global\long\def\Det{\textrm{Det}}%
\global\long\def\sgn{\textrm{sgn}}%
\global\long\def\pr{^{\prime}}%
\global\long\def\bn#1{\mbox{\boldmath\ensuremath{#1}}}%
\global\long\def\bB{\bn B}%
\global\long\def\bP{\bn P}%
\global\long\def\bY{\bn Y}%
\global\long\def\bV{\bn V}%
\global\long\def\bW{\bn W}%
\global\long\def\bG{\bn G}%
\global\long\def\bM{\bn M}%
\global\long\def\bm{\bn m}%
\global\long\def\bE{\bn E}%
\global\long\def\bK{\bn K}%
\global\long\def\bL{\bn L}%
\global\long\def\bC{\bn C}%
\global\long\def\bT{\bn T}%
\global\long\def\bg{\bn g}%
\global\long\def\bN{\bn N}%
\global\long\def\bX{\bn X}%
\global\long\def\bR{\bn R}%
\global\long\def\bHs{\bn{Hs}}%
\global\long\def\bDelta{\bn{\Delta}}%
\global\long\def\bsigma{\bn{\sigma}}%
\global\long\def\bpsi{\bn{\psi}}%
\global\long\def\bQ{\bn Q}%
\global\long\def\bZ{\bn Z}%
\global\long\def\bU{\bn U}%
\global\long\def\bz{\bn z}%
\global\long\def\bu{\bn u}%
\global\long\def\bk{\bn k}%
\global\long\def\bUpsilon{\bn{\Upsilon}}%
\global\long\def\bEta{\bn{\eta}}%
\global\long\def\bmu{\bn{\mu}}%
\global\long\def\brho{\bn{\rho}}%
\global\long\def\bpartial{\bn{\partial}}%
\global\long\def\Elecf{\mathbf{E}}%
\global\long\def\Edisplacement{\mathbf{D}}%
\global\long\def\Strain{\grad\disp}%
\global\long\def\Stress{\boldsymbol{\sigma}}%
\global\long\def\Piezoelectricmodule{\mathbf{B}}%
\global\long\def\Permittivity{\mathbf{A}}%
\global\long\def\alfacoupled{\check{\alpha}}%
\global\long\def\gammacoupled{\check{\gamma}}%
\global\long\def\bettacoupled{\check{\beta}}%
\global\long\def\rhocoupled{\check{\rho}}%
\global\long\def\Scoupled{\check{S}}%
\global\long\def\Lcoupled{\check{L}}%
\global\long\def\Qcoupled{\check{Q}}%
\global\long\def\elas{\mathbf{C}}%
\global\long\def\momentum{\mathbf{p}}%
\global\long\def\force{\mathbf{f}}%
\global\long\def\disp{\mathbf{u}}%
\global\long\def\potential{\phi}%
\global\long\def\state{\mathsf{S}}%
\global\long\def\position{\mathbf{x}}%
\global\long\def\cmat{\mathsf{C}}%
\global\long\def\rmat{\mathsf{R}}%
\global\long\def\mmat{\mathsf{m}}%
\global\long\def\pmat{\mathsf{p}}%
\global\long\def\gradt{\mathsf{B}}%
\global\long\def\genpotential{\mathsf{w}}%
\global\long\def\vol{\Omega}%
\global\long\def\rve{\Omega^{\mathrm{rve}}}%
\global\long\def\effective#1{\tilde{#1}}%
\global\long\def\ensemble#1{\left\langle #1\right\rangle }%
\global\long\def\average#1{\overline{#1}}%
\global\long\def\rg{\mathbf{W}}%
\global\long\def\po#1{#1^{\left(1\right)}}%
\global\long\def\pt#1{#1^{\left(2\right)}}%
\global\long\def\ptr#1{#1^{\left(3\right)}}%
\global\long\def\length{l}%
\global\long\def\dt#1{\dot{#1}}%
\global\long\def\willis{\mathbf{S}}%
\global\long\def\area{a}%
\global\long\def\fmat{\mathsf{f}}%
\global\long\def\lmat{\mathsf{L}}%
\global\long\def\hmat{\mathsf{h}}%
\global\long\def\bmat{\mathsf{b}}%
\global\long\def\green{\mathsf{G}}%
\global\long\def\adjoint#1{#1^{\dagger}}%
\global\long\def\cc#1{#1^{*}}%
\global\long\def\intvol#1{\int_{\vol}#1\mathrm{d}\vol}%
\global\long\def\boundaryw{\bnd_{\mathsf{w}}}%
\global\long\def\intsurfacepot#1{\int_{\boundaryw}#1\mathrm{d}\area}%
\global\long\def\boundaryt{\bnd_{\mathsf{t}}}%
\global\long\def\boundary{\bnd}%
\global\long\def\intsurfaceh#1{\int_{\boundaryt}#1\mathrm{d}\area}%
\global\long\def\normalop{\mathsf{n}}%
\global\long\def\smat{\mathsf{s}}%
\global\long\def\rene#1{#1^{rene}}%
\global\long\def\fmat{\mathsf{f}}%
\global\long\def\dmat{\mathsf{D}}%
\global\long\def\zm#1#2{\mathsf{0}_{#1\times#2}}%
\global\long\def\nmat{\mathsf{N}}%
\global\long\def\tp#1{#1^{\mathsf{T}}}%
\global\long\def\charge{q}%
\global\long\def\scharge{\omega}%
\global\long\def\dirac{\delta}%
\global\long\def\green{\mathsf{G}}%
\global\long\def\adjoint#1{#1^{\dagger}}%
\global\long\def\cc#1{#1^{*}}%
\global\long\def\intvol#1{\int_{\vol}#1\mathrm{d}\vol}%
\global\long\def\boundaryw{\bnd_{\mathsf{w}}}%
\global\long\def\intsurfacepot#1{\int_{\boundaryw}#1\mathrm{d}\area}%
\global\long\def\boundaryt{\bnd_{\mathsf{t}}}%
\global\long\def\intsurfaceh#1{\int_{\boundaryt}#1\mathrm{d}\area}%
\global\long\def\normalop{\mathsf{n}}%
\global\long\def\smat{\mathsf{s}}%
\global\long\def\rene#1{#1^{rene}}%
\global\long\def\fmat{\mathsf{f}}%
\global\long\def\dmat{\mathsf{D}}%
\global\long\def\zm#1#2{\mathsf{0}_{#1\times#2}}%
\global\long\def\nmat{\mathsf{N}}%
\global\long\def\tp#1{#1^{\mathsf{T}}}%
\global\long\def\charge{q}%
\global\long\def\scharge{\omega}%
\global\long\def\dirac{\delta}%
\global\long\def\period{l}%
\global\long\def\scalarf{f}%
\global\long\def\scalarD{D}%
\global\long\def\randomvar{p}%
\global\long\def\cell{\vol_{\mathrm{p}}}%
\global\long\def\realization{y}%
\global\long\def\realizationf{\zeta_{\realization}}%
\global\long\def\periodicf{\zeta_{\mathrm{p}}}%
\global\long\def\realizations#1{#1_{y}}%
\global\long\def\ensembleint#1{\int_{\cell}#1\mathrm{d}\realization}%
\global\long\def\periodic#1{#1_{\mathrm{p}}}%
\global\long\def\pl#1{#1^{+}}%
\global\long\def\mn#1{#1^{-}}%
\global\long\def\scalarG{G}%
\global\long\def\zerosub#1{#1_{0}}%
\global\long\def\parameter{y}%
\global\long\def\samplespace{\Lambda}%
\global\long\def\constsub#1{#1_{c}}%
\global\long\def\wronskian{V}%
\global\long\def\stressindex{\sigma}%
\global\long\def\Edisplaindex{D}%
\global\long\def\dispIndex{u}%
\global\long\def\forceindex{f}%
\global\long\def\Gindex{G}%
\global\long\def\elasindex{C}%
\global\long\def\Piezoelectricindex{B}%
\global\long\def\Permittivityindex{A}%
\global\long\def\eigenindex{\eta}%
\global\long\def\nhindex{n}%
\global\long\def\positionscalar{x}%
\global\long\def\plmn#1{#1^{\pm}}%
\global\long\def\coeffG{V}%
\global\long\def\scalarmomentum{p}%
\global\long\def\Piezoelectricmoda{\Piezoelectricmodule_{1}}%
\global\long\def\Piezoelectricmodb{\Piezoelectricmodule_{2}}%
\global\long\def\willisa{\mathcal{S}_{1}}%
\global\long\def\willisb{\mathcal{S}_{2}}%
\global\long\def\rga{\mathcal{W}_{1}}%
\global\long\def\rgb{\mathcal{W}_{2}}%
\global\long\def\greentensor{\mathbf{G}}%
\global\long\def\elascalar{C}%
\global\long\def\Piezoelectricscalar{B}%
\global\long\def\Permittivityscalar{A}%
\global\long\def\williscalar{S}%
\global\long\def\rgscalar{W}%
\global\long\def\scalarelas{C}%
\global\long\def\scalarpiezo{B}%
\global\long\def\scalarA{A}%
\global\long\def\scalarS{S}%
\global\long\def\scalarW{W}%
\global\long\def\sure{\mathsf{a}}%
\global\long\def\equivc{\check{\scalarelas}}%
\global\long\def\phase#1{#1^{\left(n\right)}}%
\global\long\def\constanta{\alpha}%
\global\long\def\constantb{\beta}%
\global\long\def\statev{\mathsf{s}}%
\global\long\def\coefficients{\varsigma}%
\global\long\def\adjointerm{\vartheta}%
\global\long\def\abreviata{\zeta}%
\global\long\def\abreviatb{\gamma}%
\global\long\def\qphase#1{#1^{\left(p\right)}}%
\global\long\def\qphase#1{#1^{\left(q\right)}}%
\global\long\def\sol{\mathcal{U}}%

\title{Symmetry Breaking Creates Electro-Momentum Coupling in Piezoelectric
Metamaterials }
\author{René Pernas-Salomón and Gal Shmuel}
\maketitle

\lyxaddress{Faculty of Mechanical Engineering, Technion--Israel Institute of
Technology, Haifa 32000, Israel}
\begin{abstract}
The momentum of deformable materials is coupled to their velocity.
Here, we show that in piezoelectric composites which deform under
electric fields, the momentum can also be coupled  to the electric
stimulus by a designed macroscopic property. To this end, we assemble
these materials in a pattern with asymmetric microstructure,  develop
a theory to calculate the relations between the macroscopic fields,
and propose a realizable system  that exhibits this coupling.
In addition to its fundamental importance, our design thus forms
a metamaterial for mechanical wave control, as traversing waves are
governed by the balance of momentum, and, in turn, the engineered
electro-momentum coupling. While introduced for piezoelectric materials,
our analysis immediately applies to piezomagnetic materials, owing
to the mathematical equivalence between their governing equations,
and we expect our framework to benefit other types of elastic media
that respond to non-mechanical stimuli. 

Keywords: dynamic homogenization, piezoelectric composite, Bloch Floquet
waves, metamaterials, effective properties, constitutive relations,
wave propagation, Willis coupling 

\end{abstract}

\section{Introduction}

The effective or macroscopic properties of materials  are  modeled
by the coupling parameters between  physical fields in germane constitutive
equations \cite{Truesdell1960hc}; extraordinary properties are
engineered by cleverly designing the microstructure of artificial
materials. 
Such metamaterials were developed in optics, acoustics and mechanics
for various objectives \cite{Wegener2013,Florijn2014prl,Meza2015pnas,Kadi2019nrp}.

One of the grand challenges in metamaterial design is to obtain control
over traversing waves \cite{liu2007PRL,Chen2010nm,banerjee2011introduction,craster2012acoustic,PARNELL2013WM,Wang2014prl,gonella15,Moleron2015nc,Cummer2016,phani2017book,deng2019prl}.
Mechanical waves are governed by the balance of momentum; at the
microscale, the momentum is coupled only to the material  point velocity
by the mass density.  Willis discovered that in elastic materials
with a specific microstructure, the macroscopic momentum is coupled
also to the strain by the now termed Willis coupling \cite{willis1997book}.
This coupling thus offers a designable degree of freedom to manipulate
waves. 

A series of theoretical studies were carried out to characterize Willis
coupling and understand its physical origins \cite{nassar2015willis,XIANG2016JMPS,NASSAR2016prsa,Muhlestein20160Prsa2,NASSAR2017jmps,Sieck2017prb,su2018prb,quan2018prl,Meng2018prsa}.
Guided by accompanying predictions that Willis coupling is connected
with unusual phenomena such as asymmetric reflections and unidirectional
transmission, recent experimental realizations of Willis metamaterials
that demonstrate these phenomena were reported \cite{Koo2016nc,Muhlestein2017nc,li2018nc,Popa2018nc,Yao20182IJSS,Merkel2018prb,Melnikov2019nc,Liu2019prx,zhai2019arxiv}.
To date, these investigations were limited to metamaterials that are
deformable only by mechanical forces. 

\begin{figure}[t!]
\centering 
\includegraphics[width=0.50\textwidth]{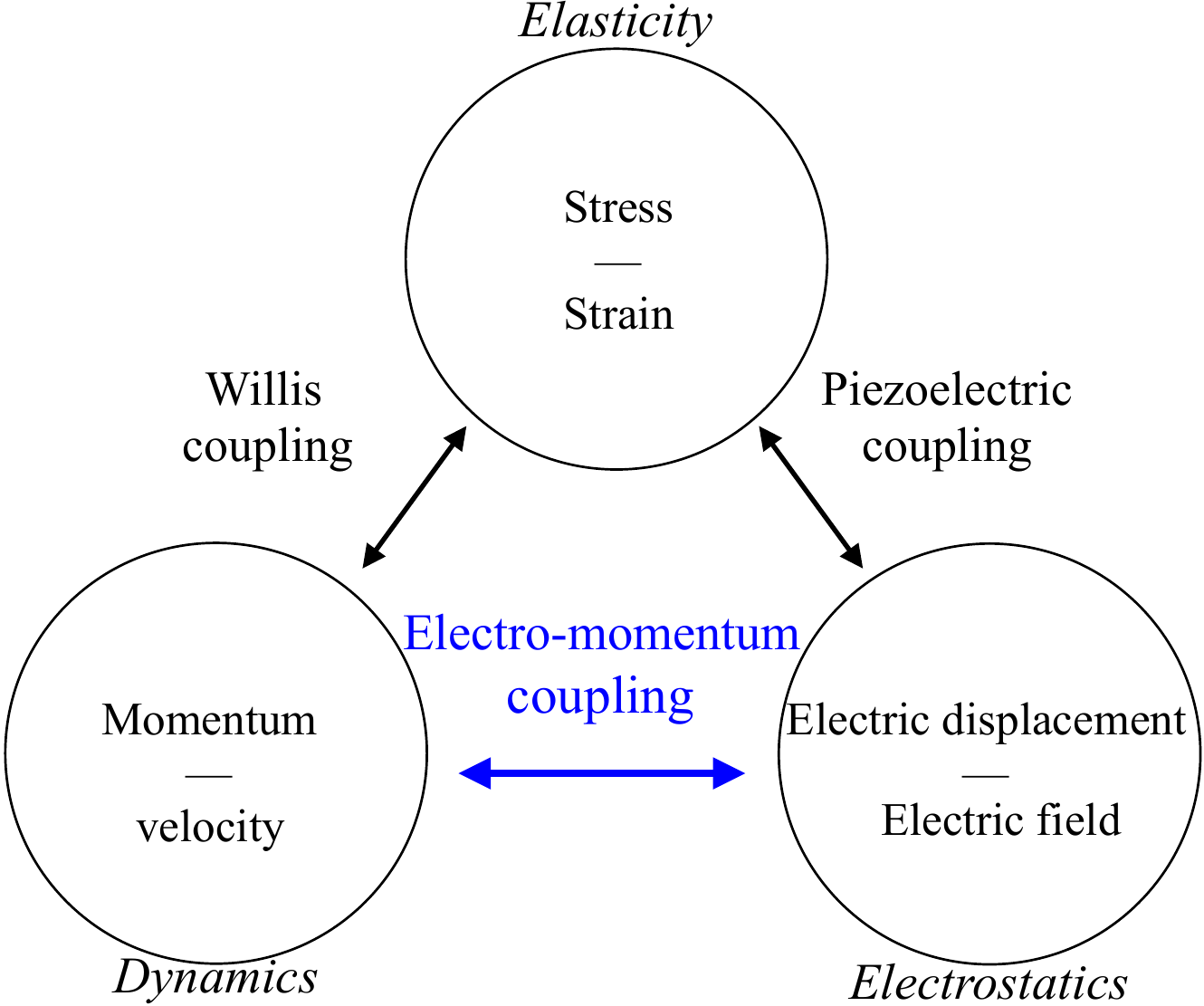} \caption{\label{fig:coupling}Schematics of   the   electro-momentum coupling reported here,  with respect to the intrinsic piezoelectric coupling and Willis metamaterial coupling. Similarly to   piezoelectric and Willis couplings, the electro-momentum coupling appears in (meta)materials with no inversion symmetry. While the diagram refers to elasticity and electrostatics, it applies to other branches of physics: electrostatics is replaceable by magnetostatics,  and elasticity by fluid mechanics.}
\end{figure}

Here, we consider constituents that additionally deform by non-mechanical
stimuli, namely, piezoelectric materials responding to electric fields
\cite{Mason1950fk,Zelisko2014nc}. We construct asymmetric patterns
of such responsive materials, and show that their macroscopic momentum
can additionally be coupled by design to the stimulus. We call this
macroscopic property the electro-momentum coupling (Fig.~\ref{fig:coupling}).
Akin to the intrinsic piezoelectric and engineered Willis couplings,
this coupling appears in (meta)materials with no inversion symmetry.
Beyond its theoretical significance, mechanical metamaterials designed
with this property can actively manipulate waves by modulation of
the external stimulus, contrary to typical metamaterials whose functionality
is fixed, cf.~Refs.~ \cite{Lapine2011nm,Cha2018nature,Jackson2018sciadv,SHI2019acta}.
To this end, it is required to carry out complementary studies on
the connection between the electro-momentum coupling, scattering properties,
and medium composition \cite{pernas2019b}, similarly to the process
that was required in employing Willis coupling for metamaterial design
\cite{Muhlestein2017nc,quan2018prl,Liu2019prx}; the present work
opens the route for these studies. 

The macroscopic properties of metamaterials are analytically calculated
using homogenization or effective medium theories \cite{Smith2006JOSA,FIETZ2010pysicaaB,alu2011prb,Shuvalov2011prsa,nemat2011homogenization,Torrent2011njp,Antonakakis2013PRSA,srivastava2015elastic,Muhlestein2016prsa1,Ponge2017EML}.
Guided by the effective elastodynamic theory of Willis \cite{Willis1981WM,willis1981variational,Willis1985IJSS,Willis2009,Willis2011PRSA,WILLIS2012MOM,WILLIS2012MRC},
we develop a homogenization method for piezoelectric metamaterials,
whose application unveils the electro-momentum coupling. Our method
is based on three elements. Firstly, it employs a unified framework
we developed to account for the microscopic interactions between the
mechanical and non-mechanical fields. Secondly, it uses an averaging
scheme whose resultant effective fields identically satisfy the macroscopic
governing equations. To this end, we have adapted the ensemble averaging
approach of Willis \cite{Willis2011PRSA} to the current setting.
Lastly, it incorporates driving forces that render the effective properties
unique, as firstly advocated by Fietz and Shvets \cite{FIETZ2010pysicaaB},
and later in Refs.~\cite{Willis2011PRSA,Sieck2017prb}.

Before proceeding, a short discussion on the applicability of Willis
homogenization scheme, and by transitivity our scheme, is in order.
 First, we note that the scheme is independent of any assumptions,
and delivers effective properties that identically satisfy the field
equations and boundary conditions, therefore considered exact \cite{nassar2015willis,Meng2018prsa};
in fact, asymptotic homogenization schemes were shown to be approximations
of Willis homogenization \cite{NASSAR2016prsa,Meng2018prsa}\textit{\emph{.}}
Thus, for infinite periodic medium, the scheme reproduces precisely
the its corresponding band diagram. Still, for the homogenized fields
to serve as good approximations of the microscopic ones, certain homogenizability
conditions should be satisfied \cite{nassar2015willis}. These show
that beyond the long-wavelength low-frequency limit there is an additional
range in which dynamic homogenization is meaningful \cite{srivastava2015elastic}.\textit{\emph{
We further note that the new coupling terms emerging from this scheme
were found essential for obtaining an effective description that satisfies
fundamental principles such as causality \cite{Sieck2017prb}, similarly
to the}} need for the analogous bianisotropic coupling in electromagnetics
\cite{Alu2011-PhysRevB,alu2011prb}. 

The paper is structured as follows. In Sec.$\,$\ref{sec:Equations-of-responsive}
we summarize the equations governing elastic waves in heterogeneous
linearly responsive media, \emph{i.e.}, media that respond mechanically
to non-mechanical loads. In Sec.$\,$\ref{sec:Derivation-of-the}
we develop our dynamic homogenization scheme for such media, and highlight
the essential features of the resultant macroscopic properties. We
apply our theory to longitudinal waves in piezoelectric layers assembled
periodically with broken inversion symmetry in Sec.$\,$\ref{sec:Application-to-piezoelectric}.
A summary of our work concludes the paper in Sec.$\,$\ref{sec:Summary}. 

Before we proceed, we stress out that while this introduction was
concerned with piezoelectric materials which respond to electric stimuli,
our results immediately apply to piezomagnetic materials, owing to
the mathematical similarity between their microscopic equations. We
also expect our framework to benefit other types of elastic media
that respond to non-mechanical stimuli, such as thermoelastic composites.
These extensions are discussed in Appendix A

\section{\label{sec:Equations-of-responsive}Equations of piezoelectric composites}

Consider a linear material with mass density $\rho$ and stiffness
tensor $\elas$ occupying the volume $\vol$, subjected to prescribed
body force density $\force$ and inelastic strain $\eigen$. Physically,
$\eigen$ can result from a plastic process or phase transformation,
where the mathematical motivation to account for it will be explained
in the sequel \cite{FIETZ2010pysicaaB,Willis2011PRSA}.  The response
of the material is governed by the balance of linear momentum $\momentum$
\begin{eqnarray}
\begin{aligned}\div{\Stress}+\force-\dt{\momentum}=\mathbf{0},\end{aligned}
\label{NewtonLaw}
\end{eqnarray}
where $\Stress$ is the Cauchy stress second-order tensor and the
superposed dot denotes time derivative. The material mechanically
interacts with the electric displacement $\Edisplacement$, and electric
field $\Elecf$, which satisfy 
\begin{eqnarray}
 & \grad\cdot\Edisplacement=\charge,\quad\curl\times\Elecf=\mathbf{0},\label{GaussLaw}
\end{eqnarray}
where $\charge$ is a prescribed free charge density, acting as a
source similarly to $\force$. Note that Eq.~\eqref{GaussLaw}$_{2}$
is identically satisfied by setting $\Elecf=-\grad\potential$. 

The constitutive relations between the fields are written in symbolic
matrix form \cite{auld1973acoustic}
\begin{equation}
\left(\begin{array}{c}
\Stress\\
\Edisplacement\\
\momentum
\end{array}\right)=\left(\begin{array}{ccc}
\elas & \tp{\Piezoelectricmodule} & 0\\
\Piezoelectricmodule & -\Permittivity & 0\\
0 & 0 & \rho
\end{array}\right)\left(\begin{array}{c}
\grad\disp-\eigen\\
\grad\potential\\
\dt{\disp}
\end{array}\right),\label{ConstiRel-1}
\end{equation}
where $\disp$ is the displacement field, $\Permittivity$ and $\Piezoelectricmodule$
are the dielectric and piezoelectric (tensorial) properties, and the
transpose of $\Piezoelectricmodule$ is defined by $\tp B_{ijk}=\scalarpiezo_{kij}$.
We assume the standard tensor symmetries, which in components read
\begin{equation}
\scalarA_{ij}=\scalarA_{ji},\scalarpiezo_{ijk}=\scalarpiezo_{jik},\scalarelas_{ijkl}=\scalarelas_{jikl}=\scalarelas_{jilk}=\scalarelas_{klij},\sigma_{ij}=\sigma_{ji}.\label{eq:symmetries}
\end{equation}
For later use, we denote the matrix (resp.~column vector) in the
right (resp.~left) hand side of Eq.$\,$\eqref{ConstiRel-1} by $\lmat$
(resp.~$\hmat$). 

We clarify that the elements in the symbolic matrices here and in
what follows are differential operators and tensors of different order,
and their product should be interpreted accordingly. For example,
the product $\lmat_{11}\mmat_{1}$ represents double contraction,
which in tensor notation is $\elas:\eigen$, and in index notation
is $C_{ijkl}\eta_{kl}$, while $\lmat_{12}\bmat_{2}$ is the single
contraction $\tp{\Piezoelectricmodule}\cdot\grad\potential$, or $\tp B_{ijk}\potential_{,k}$.
We further note that the symbolic matrix structure can be cast into
standard matrix representation, using Voigt notation. Accordingly,
$\elas,\tp{\Piezoelectricmodule}$ and $\Piezoelectricmodule$ are
representable by $6\times6,6\times3$ and $3\times6$ matrices, respectively,
such that $\lmat$ is a \emph{symmetric} $12\times12$ matrix. (The
zeros are $9\times3$ and $3\times9$ null matrices, and $\rho\identity$
is a $3\times3$ matrix.) The symmetric tensor $\Stress$ is mapped
to a $6\times1$ column vector (and so is the symmetric part of $\grad\disp$),
such that $\hmat$ is a $12\times1$ column vector, and so on. 

The prescribed boundary conditions are $\disp=\disp_{0}$ and $\potential=\potential_{0}$
over $\boundaryw\subset\bnd$, and across the remaining boundary $\boundaryt=\bnd\backslash\boundaryw$
are $\Stress\cdot\nh=\traction_{0}$ and $\Edisplacement\cdot\nh=-\surfacech$,
where $\nh$ is the outward normal, $\traction_{0}$ is the traction,
and $\surfacech$ is the surface charge density. 

When the medium is randomly heterogeneous, its properties $\rho,\Permittivity,\Piezoelectricmodule$
and $\elas$ are functions of the position $\position$ and a parameter
$\parameter$ of a sample space $\samplespace$ with certain probability
measure. Importantly, a periodic medium---the prevalent case of interest
for metamaterials---can be analyzed as random, by considering different
realizations of the composite generated by periodizing representative
volume elements whose corner is a uniformly distributed random variable,
and identifying $y$ with this variable. \cite{Willis2011PRSA,WILLIS2012MRC}. 

\begin{figure}[t!]
\centering \includegraphics[width=0.45\textwidth]{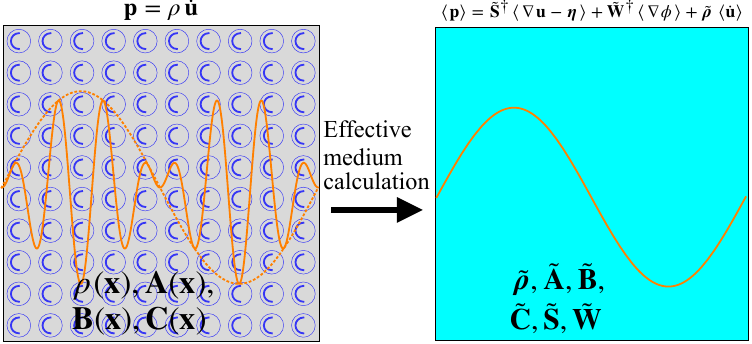}
\caption{\label{fig:homogenization}Piezoelectric composite with asymmetric
periodic cell, subjected to independent external sources. At the
microscale, its momentum is $\protect\momentum=\rho\protect\dt{\protect\disp}$.
Our effective medium theory reveals that the macroscopic momentum
is $\protect\ensemble{\protect\momentum}=\protect\adjoint{\protect\effective{\protect\willis}}\protect\ensemble{\protect\grad\protect\disp-\protect\eigen}+\protect\adjoint{\protect\effective{\protect\rg}}\protect\ensemble{\protect\grad\protect\potential}+\protect\effectivemass\protect\ensemble{\protect\dt{\protect\disp}}$.}
\end{figure}

Next, observe that ensemble averaging of Eqs.~\eqref{NewtonLaw}
and \eqref{GaussLaw} over $\parameter\in\samplespace$, denoted by
$\ensemble{\cdot}$, provides 
\begin{equation}
\begin{aligned}\grad\cdot\ensemble{\Stress}+\force-\dt{\ensemble{\momentum}}=\mathbf{0},\end{aligned}
\quad\grad\cdot\ensemble{\Edisplacement}=\charge,\label{ensembledeqs}
\end{equation}
where $\ensemble{\force}=\force$ and $\ensemble q=\charge$ since
$\force$ and $\charge$ are sure (prescribed). Eq.~\eqref{ensembledeqs}
suggests the use of $\ensemble{\Stress},\ensemble{\Edisplacement}$
and $\ensemble{\momentum}$ as effective fields that identically satisfy
the governing equations. The effective properties are thus the quantities
that relate these effective fields with $\ensemble{\grad\disp},\ensemble{\dt{\disp}}$
and $\ensemble{\grad\potential}$, to form effective constitutive
relations.  Together with Eq.~\eqref{ensembledeqs}, they establish
a meaningful description of the material when the ensemble averaged
fields fluctuate slowly enough relatively to the scale of the microstructure;
for a rigorous description of the applicability conditions for homogenization
see Ref.~\cite{nassar2015willis}. In periodic media undergoing Bloch-Floquet
waves, these ensemble averages reduce to volume averages over the
periodic part of each field\footnote{The equivalence between ensemble averaging and volume averaging of
the periodic part will be exemplified in Sec.$\ $\ref{sec:Application-to-piezoelectric}
in the scalar case, without the loss of generality. } \cite{Willis2011PRSA,nassar2015willis}. Qualitatively, the effective
fields have the form of the curve in the right sketch of Fig.$\,$\ref{fig:homogenization},
after the fluctuations of the curve in the left sketch have been averaged
out. The outstanding problem is to calculate the effective properties.
Before we derive them, we can now provide a formal statement of our
main result: homogenization shows that the effective constitutive
relations are in the form 
\begin{eqnarray}
\left(\begin{array}{c}
\ensemble{\Stress}\\
\ensemble{\Edisplacement}\\
\ensemble{\momentum}
\end{array}\right)=\left(\begin{array}{ccc}
\effective{\elas} & \tp{\tilde{\Piezoelectricmodule}} & \effective{\willis}\\
\effective{\Piezoelectricmodule} & -\tilde{\Permittivity} & \effective{\rg}\\
\adjoint{\effective{\willis}} & \adjoint{\effective{\rg}} & \effectivemass
\end{array}\right)\left(\begin{array}{c}
\ensemble{\grad\disp}-\eigen\\
\ensemble{\grad\potential}\\
\ensemble{\dt{\disp}}
\end{array}\right),\label{Eff-constR-final}
\end{eqnarray}
with  the electro-momentum coupling $\adjoint{\effective{\rg}}$
(Fig.~\ref{fig:homogenization}), where $\adjoint{\left(\cdot\right)}$
denotes the adjoint operator with respect to the spatial variable.
Thus, our homogenization process exposes effective couplings between
$\ensemble{\Edisplacement}$ and the average velocity, and between
$\ensemble{\grad\potential}$ and the average momentum. We denote
the matrix of the effective properties in Eq.~\eqref{Eff-constR-final}
by $\effective{\lmat}$. The terms $\ensemble{\grad\disp}-\eigen$
and $\ensemble{\dt{\disp}}$ that $\effective{\lmat}$ operates on
are independent, owing to $\eigen$, thus rendering $\effective{\lmat}$
unique; otherwise, the fields $\ensemble{\grad\disp}$ and $\ensemble{\dt{\disp}}$
are derived from the same potential, resulting with non-unique $\effective{\lmat}$
\cite{Willis2011PRSA}, see, \emph{e.g.}, the example in Ref.$\ $\cite{pernassalomon2018jmps}.
The calculation of $\effective{\lmat}$ is detailed next. 

\section{\label{sec:Derivation-of-the}Derivation of the effective properties}

We adapt the ingenious approach of Willis \cite{Willis2011PRSA} to
responsive metamaterials as follows. Firstly, we cast the problem
into matrix equations whose entries are tensors and other operators.
Accordingly, Eqs.~\eqref{NewtonLaw}, \eqref{GaussLaw}$_{1}$ and
\eqref{ConstiRel-1} take the form 

\begin{eqnarray}
&&\dmat^{\mathsf{T}}\hmat=-\fmat,\label{Governing-Eqs-Willis-form}\\
&&\hmat=\lmat\left(\bmat-\mmat\right),
\label{ConstiRel-1-matrix}
\end{eqnarray}where 

\begin{eqnarray}
\begin{aligned}
&\gradt=\left(\begin{array}{cc}\grad & 0\\0 & \grad\\s & 0\end{array}\right),
\mmat=\left(\begin{array}{c}\eigen\\0\\0\end{array}\right),\dmat=\left(\begin{array}{cc}\div{} & 0\\0 & \div{}\\-s & 0\end{array}\right),\\
&\genpotential=\left(\begin{array}{c}\disp\\\potential\end{array}\right),
\fmat=\left(\begin{array}{c}\force\\-\charge\end{array}\right),\bmat=\gradt\genpotential,
\nonumber
\end{aligned}
\end{eqnarray} and we employed the Laplace transform to replace time derivatives
by products with $s$; to reduce notation, we retain the same symbols
for the temporal and transformed fields, bearing in mind they now
represent functions of $s$. 

We define the Green's function $\green\left(\position,\position'\right)$
via 
\begin{equation}
\dmat^{\mathsf{T}}\lmat\gradt\green=-\left(\begin{array}{cc}
\dirac\left(\position-\position'\right)\identity & 0\\
0 & \dirac\left(\position-\position'\right)
\end{array}\right),\label{eq:green equation}
\end{equation}
where $\identity$ is the second order identity tensor and $\dirac$
is the Dirac delta; the entries $\green_{11}$, $\green_{12}$ and
$\green_{21}$, and $\green_{22}$ are second order tensor-, vector-,
and scalar-valued functions, respectively, with the homogeneous boundary
conditions
\begin{equation}
\green=0\ \mathrm{over}\ \boundaryw,\quad\tp{\left(\gradt\green\right)}\lmat\nmat=0\ \mathrm{over}\ \boundaryt,\label{eq:bcG}
\end{equation}
where
\[
\tp{\nmat}=\left(\begin{array}{ccc}
\otimes\nh & 0 & 0\\
0 & \otimes\nh & 0
\end{array}\right).
\]
We derive next a useful expression for $\genpotential$, by right
multiplying the transpose of Eq.~\eqref{eq:green equation} by $\genpotential\left(\position\right)$,
subtracting it from the left product of Eq.~\eqref{Governing-Eqs-Willis-form}
by $\tp{\green}\left(\position,\position'\right)$, and integrating
the difference over the volume $\position'\in\vol$. The result is

\begin{eqnarray}\label{potentialG} 
\begin{aligned}
\genpotential(\position,\parameter)&=\intvol{\green^{\mathsf{T}}\fmat}+\intsurfaceh{\green^{\mathsf{T}}\tp{\nmat}\lmat\left(\gradt\genpotential-\mmat\right)}\\
&+\intvol{\left(\gradt\green\right)^{\mathsf{T}}\lmat\mmat}-\intsurfacepot{\left(\gradt\green\right)^{\mathsf{T}}\lmat\nmat\genpotential} ,
\end{aligned}
\end{eqnarray}The development of Eq.~\eqref{potentialG} employs the divergence
theorem, boundary conditions for $\green$ and symmetries of $\Permittivity,\Piezoelectricmodule$
and $\elas$, and its detailed derivation using index and tensor
notations is provided in Appendix B. Next, we manipulate the last
integral in Eq.~\eqref{potentialG} as follows. First, we formally
extend the integral domain from $\boundaryw$ to the whole boundary
$\bnd$ using the fact that the integrand vanishes over $\boundaryt$,
owing to homogeneous boundary conditions that $\green$ satisfies
over $\boundaryt$. Next, we can replace $\genpotential$ with $\ensemble{\genpotential}$,
since it is sure on $\boundaryw$ and the integrand vanishes where
it is not. By transforming the surface integral it back to the volume
via the divergence theorem, and with the aid of Eq.~\eqref{eq:green equation},
we can rewrite $\genpotential$ as \begin{eqnarray}\label{w2} 
\begin{aligned}
\genpotential(\position,\parameter)&=\intvol{\green^{\mathsf{T}}\fmat}+\intsurfaceh{\green^{\mathsf{T}}\tp{\nmat}\lmat\left(\gradt\genpotential-\mmat\right)}\\
&-\intvol{\left(\gradt\green\right)^{\mathsf{T}}\lmat\left(\ensemble{\gradt\genpotential}-\mmat\right)}+\ensemble{\genpotential}.
\end{aligned}
\end{eqnarray}Using the ensemble average of Eq.~\eqref{w2} and the fact that $\mmat,\fmat,\traction_{0}$
and $\surfacech$ are sure, we have that (see Appendix B for more
details)
\begin{equation}
\genpotential=\left\{ \tp{\green}\ensemble{\green}^{-\mathsf{T}}\ensemble{\tp{\left(\gradt\green\right)}\lmat}-\tp{\left(\gradt\green\right)}\lmat\right\} \left(\ensemble{\gradt\genpotential}-\mmat\right)+\ensemble{\genpotential}.\label{eq:wensemble}
\end{equation}
Finally, we substitute Eq.~\eqref{eq:wensemble} into Eq.~\eqref{ConstiRel-1-matrix}
and ensemble average the result to obtain the effective constitutive
relations
\begin{equation}
\ensemble{\hmat}\left(\position\right)=\effective{\lmat}\left(\ensemble{\gradt\genpotential}-\mmat\right),\label{eq:effective relation}
\end{equation}
with 

\begin{equation}
\effective{\lmat}=\ensemble{\lmat}-\ensemble{\lmat\gradt\tp{\left(\gradt\green\right)}\lmat}+\ensemble{\lmat\gradt\tp{\green}}\ensemble{\green}^{-\mathsf{T}}\ensemble{\tp{\left(\gradt\green\right)}\lmat}.\label{Eff-L}
\end{equation}

\noindent Eq.~\eqref{Eff-L} generalizes the result of Willis \cite{Willis2011PRSA}
to metamaterials that interact with non-mechanical fields. The components
of $\effective{\lmat}$ define the effective properties in Eq.~\eqref{Eff-constR-final},
which are non-local operators in space  and time. The non-zero adjoint
terms $\effective{\lmat}_{23}$ and $\effective{\lmat}_{32}$ reveal
the coupling between $\ensemble{\Edisplacement}$ and $\ensemble{\dt{\disp}}$,
and between $\ensemble{\momentum}$ and $\ensemble{\grad\potential}$,
respectively, denoted by $\effective{\rg}$. In the case that $\Edisplacement$
and $\grad\potential$ are vectors, the coupling $\effective{\rg}$
is a second order tensor. Owing to the symmetries of $\Permittivity,\Piezoelectricmodule$
and $\elas$, the operator $\lmat$ is symmetric, so that $\green$
self-adjoint with the usual symmetries of Green functions, and hence
$\effective{\lmat}$ is self-adjoint as well, justifying the notation
$\adjoint{\left(\cdot\right)}$ for $\effective{\lmat}_{31}$ and
$\effective{\lmat}_{32}$. 

\section{\label{sec:Application-to-piezoelectric}Application to piezoelectric
layers}

We exemplify the emergence of the electro-momentum coupling by calculating
the effective properties of an infinite repetition of commercially
available piezoelectric layers. Specifically, we will study two different
periodic cells, namely, (\emph{i}) all-piezoelectric cell made of
PZT4-BaTiO$_{3}$-PVDF layers, henceforth called composition 1, and
(\emph{ii}) one piezoelectric BaTiO$_{3}$ layer between elastic Al$_{2}$O$_{3}$
layer and another elastic PMMA layer, henceforth called composition
2. The material properties of the comprising phases are given in Tab.~\ref{Table-1},
where the values for the piezoelectric materials correspond to the
coefficients in the direction of the poling.

\noindent 
\begin{table}[t!]
\centering %
\begin{tabular}{lcccc}
\hline 
Phase \; \;  & $\elascalar\,\mathrm{(GPa})$ \; \;  & $\rho\,\mathrm{(kg/m^{3}})$ \; \;  & $\Piezoelectricscalar\,\mathrm{(C/m^{2}})$ \; \;  & $\Permittivityscalar$$(\textrm{nF/m})$ \; \;\tabularnewline
\hline 
$\textrm{PZT4}$ \; \;  & $115$  & 7500  & 15.1  & 5.6 \tabularnewline
$\mathit{\textrm{BaTiO}_{\textrm{3}}}$ \; \;  & $165$  & 6020  & 3.64  & 0.97\tabularnewline
$\textrm{PVDF}$ \; \;  & 12  & 1780  & -0.027  & $0.067$\tabularnewline
$\mathit{\textrm{Al}_{\mathrm{2}}}\textrm{O}_{3}$ \; \;  & 300  & 3720  & 0  & $0.079$\tabularnewline
$\textrm{PMMA}$ \; \;  & 3.3  & 1188  & 0  & $0.023$\tabularnewline
\hline 
\end{tabular}\caption{\label{Table-1} Physical properties of the phases comprising the
periodic piezoelectric laminate. }
\end{table}

The periodic cell is denoted $\cell$, and its period is denoted $\period$;
in the calculations that follow we fix $\period=3\,$mm. The layers
are oriented such that the poling direction is along the direction
of lamination, say $x$. The composite is driven by a body force density
$f$ acting in the $x$ direction, along which axial inelastic strain
$\eta$ is present; there are no free or surface charge sources in
the problem ($\surfacech=\charge=0$). As a result, longitudinal waves
propagate in the $x$ direction, such that the problem is one-dimensional
and the pertinent fields can be treated as scalars. The objective
is to obtain the macroscopic description of this problem by means
of our homogenization scheme. 

To this end, we analyze the periodic medium as random by treating
the position of the period as a uniformly distributed random variable,
with uniform probability density $\period^{-1}$ over $\cell$. Accordingly,
any $\period$-periodic function $\zeta_{y}\left(x\right)$ in realization
$\realization\in\cell$ is $\zeta_{0}\left(x-y\right)$; its ensemble
average 
\begin{equation}
\ensemble{\zeta}=\frac{1}{\period}\ensembleint{\zeta_{0}\left(x-\realization\right)}\label{eq:ensembley}
\end{equation}
is independent of $x$, and equals the spatial average in any realization.

 We examine first the governing equations in realization $\realization=0$.
In the absence of free charge, Gauss law within any layer reads
\begin{equation}
\scalarD_{0,x}^{\left(n\right)}=0,\label{eq:gauss0}
\end{equation}
where superscript $n=a,b$ and $c$ denotes values in the first, middle,
and third layer, respectively. Note that $D$ is continuous across
the layers, and since there are no electrodes and surface charge,
we have that $D=0$ everywhere. Further, Eq.$\ $\eqref{eq:gauss0}
together with the constitutive relation 
\begin{equation}
D_{0}^{\left(n\right)}=B_{0}^{\left(n\right)}u_{0,x}^{\left(n\right)}-A_{0}^{\left(n\right)}\text{\ensuremath{\phi_{0,x}^{\left(n\right)}}}\label{eq:localD}
\end{equation}
implies that 
\begin{equation}
\text{\ensuremath{\phi_{0,xx}^{\left(n\right)}}}=\frac{B_{0}^{\left(n\right)}}{A_{0}^{\left(n\right)}}u_{0,xx}^{\left(n\right)}.\label{eq:localphi}
\end{equation}
 The second governing equation is the equation of motion, namely,
\begin{equation}
\sigma_{0,x}^{\left(n\right)}-s^{2}\rho_{0}^{\left(n\right)}u_{0}^{\left(n\right)}=-f_{0}^{\left(n\right)},\label{eq:localcauchy}
\end{equation}
and we recall that the Laplace transform has been used, as in Sec.$\ $\ref{sec:Derivation-of-the}.
Substituting in the constitutive relation for the stress
\begin{equation}
\sigma_{0}^{\left(n\right)}=C_{0}^{\left(n\right)}\left(u_{0,x}^{\left(n\right)}-\eta_{0}^{\left(n\right)}\right)+B_{0}^{\left(n\right)}\phi_{0,x}^{\left(n\right)}\label{eq:stress0}
\end{equation}
and Eq.$\ $\eqref{eq:localphi} yields 
\begin{equation}
\left(\zerosub{\scalarelas}+\dfrac{\zerosub{\scalarpiezo}^{2}}{\zerosub{\scalarA}}\right)\left(u_{0,xx}-\eta_{,x}\right)-s^{2}\zerosub{\rho}u_{0}=-\scalarf_{0},\label{eq:simplified eom}
\end{equation}
where the superscript notation was suppressed for brevity. Since the
homogeneous equation derived from Eq.~\eqref{eq:simplified eom}
has periodic coefficients, its Green function is constructed using
Bloch-Floquet solutions. The Green function and its ensemble average
are thus \begin{eqnarray}
\begin{aligned}
\zerosub G\left(x,x^{\prime}\right)&=\left\lbrace \begin{array}{l}\wronskian\pl u\left(x\right)\mn u(x^{\prime}),\quad x<x^{\prime},\\[10pt]\wronskian\pl u(x^{\prime})\mn u(x),\ \quad x^{\prime}<x,\end{array}\right.\label{GreenF}\\
\ensemble{\scalarG}\left(x,x'\right)&=\frac{1}{\period}\ensembleint{\zerosub{\scalarG}\left(x-\realization,x'-\realization\right)},\label{ensenbleG}
\end{aligned}
\end{eqnarray} where $\ensemble{\green}$ is only a function of $x-x'$, $\wronskian$
is 
\begin{equation}
\wronskian^{-1}=\left(\zerosub{\scalarelas}+\dfrac{\zerosub{\scalarpiezo}^{2}}{\zerosub{\scalarA}}\right)\left(\pl u_{,x}\mn u-\pl u\mn u_{,x}\right),\ u^{\pm}=\periodic u^{\pm}\left(x\right)e^{\pm i\kb x},
\end{equation}
and $\periodic u^{\pm}$ are $\period$-periodic functions whose standard
(and tedious) calculation is detailed in Appendix C.

As the simplicity of the problem enabled a solution via a single Green
function, a simpler equivalent to Eq.~\eqref{eq:wensemble} for $u$
follows, namely,

\begin{eqnarray}
\begin{aligned}u\left(x,y\right)= & \left(G\ensemble G{}^{-1}\ensemble{G_{,x^{\prime}}\check{\scalarelas}}-G{}_{,x^{\prime}}\check{\scalarelas}\right)\left(\ensemble{u_{,x'}}-\eta\right)\\
 & +\left(G\ensemble G{}^{-1}\ensemble{G\rho}-G\rho\right)s^{2}\ensemble u+\ensemble u,
\end{aligned}
\label{u-WillisFormulation}
\end{eqnarray}
where $\check{\scalarelas}=\scalarelas+\dfrac{\scalarpiezo^{2}}{\scalarA}$.
Observing that in the prescribed settings $\scalarD=0$ in any realization,
we obtain the remaining fields $\potential\left(x,\realization\right)$
from Eq.~\eqref{ConstiRel-1} in terms of $u\left(x,y\right),\scalarA\left(x,\realization\right)$
and $\scalarpiezo\left(x,\realization\right)$. \begin{figure}[t!]	
\centering 
\captionsetup[subfloat]{position=top} 
\subfloat[$\xi\period=0$]{\label{fig:xi0}		
\includegraphics[width=0.335\textwidth]{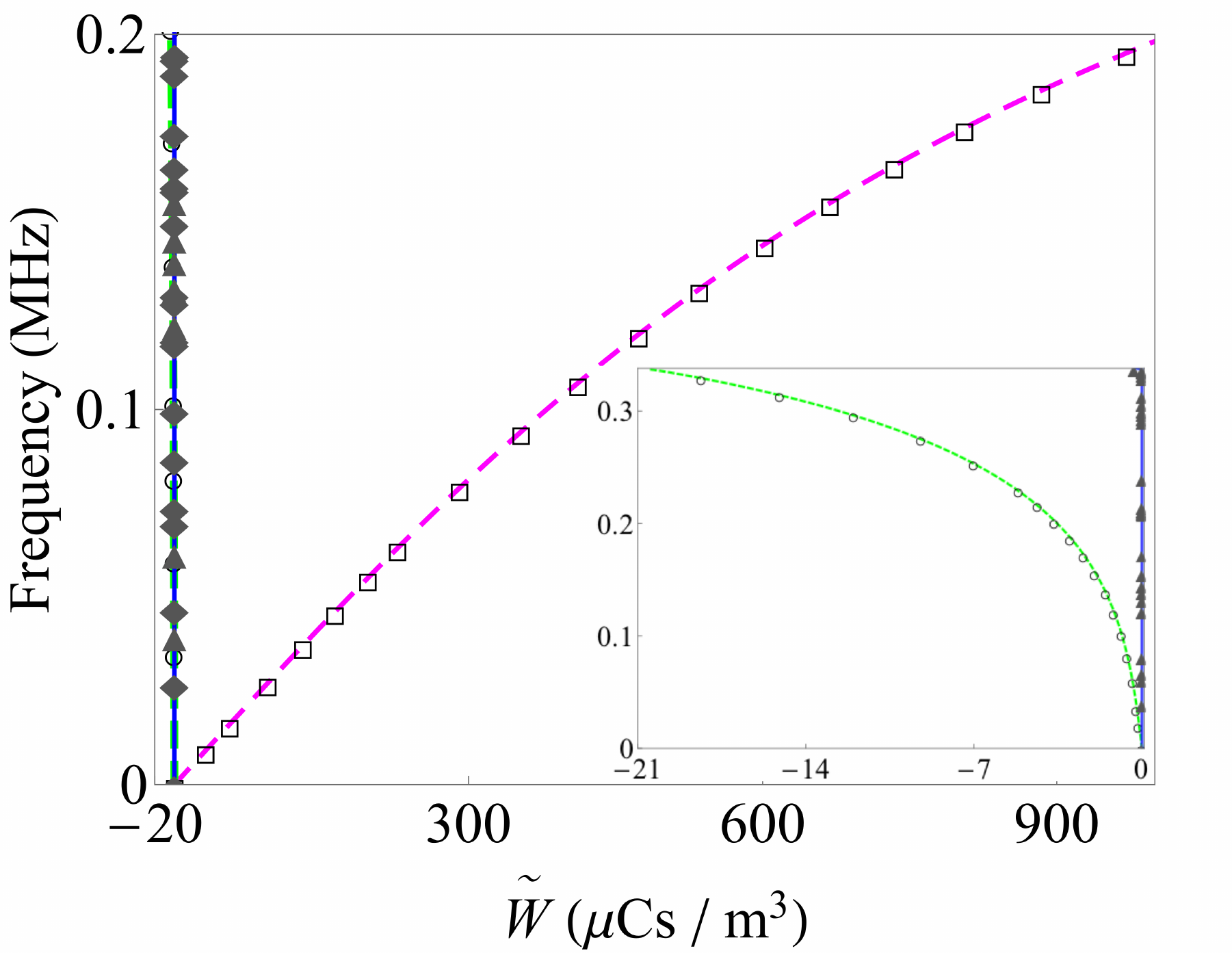}}
 \subfloat[$\xi\period=2$]{\label{fig:xi2}
\includegraphics[width=0.323\textwidth]{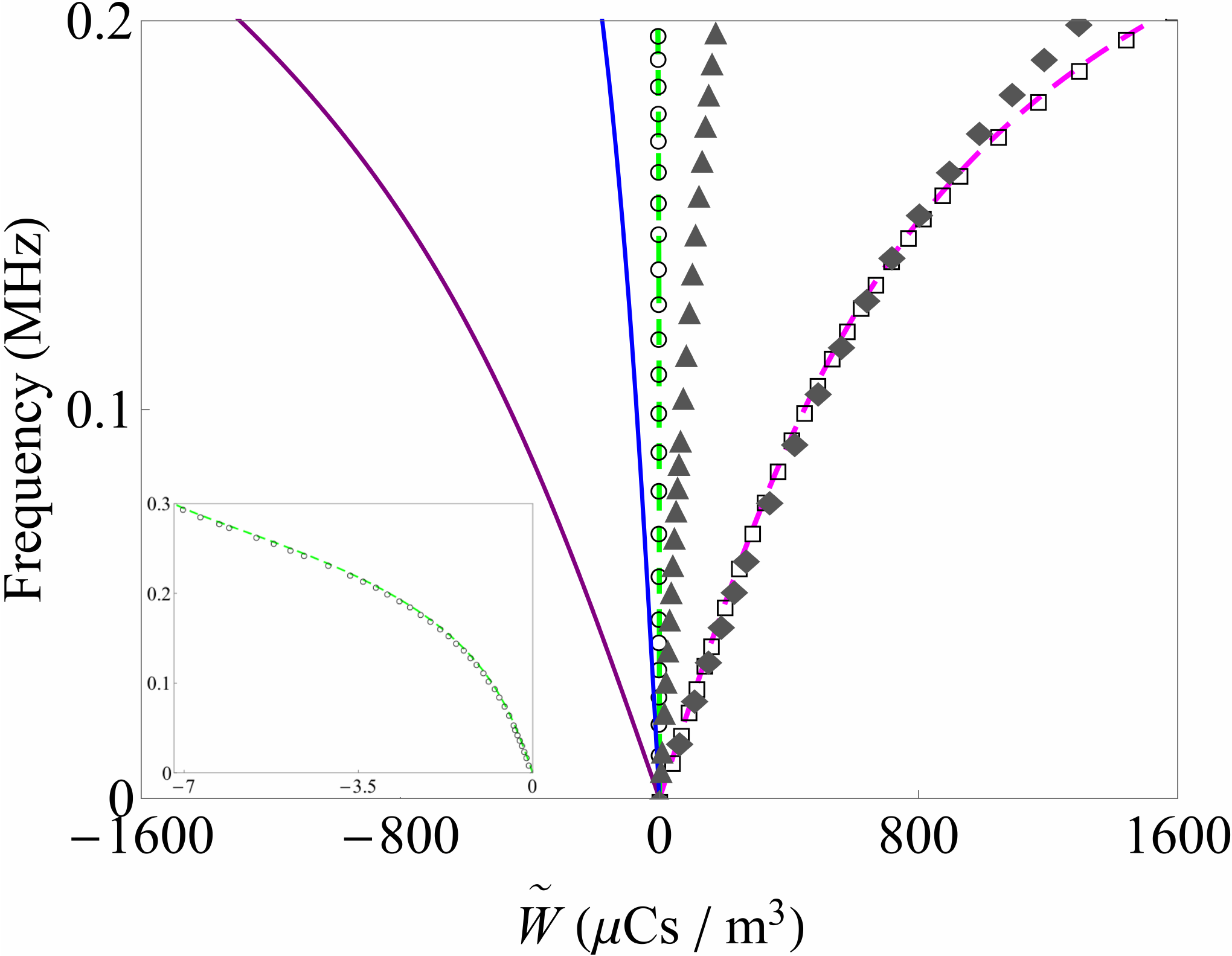}}
\subfloat[$\xi\period=2$, $\period^{(b)}=0$]{\label{fig:xi2sym}		
\includegraphics[width=0.322\textwidth]{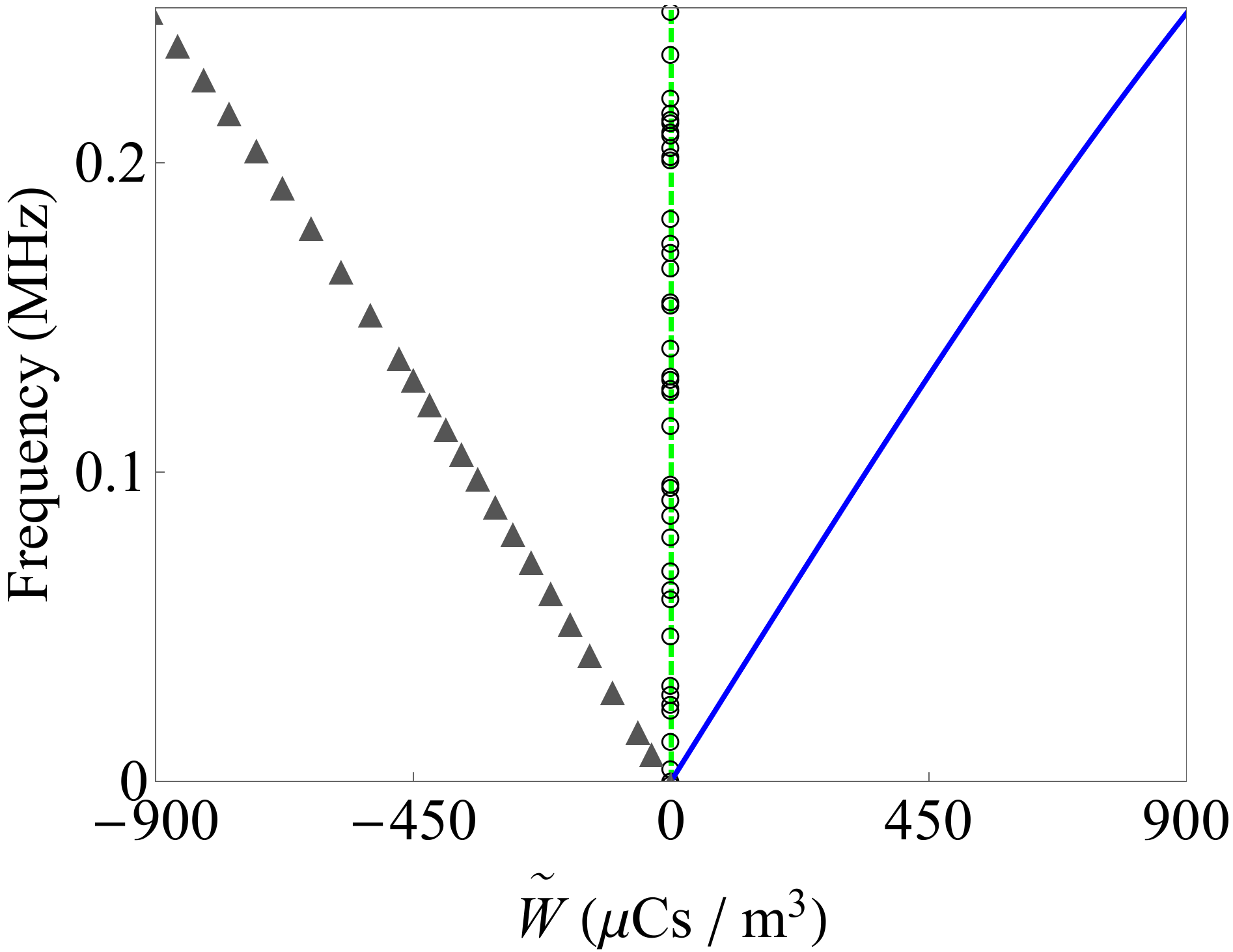}}
	\caption{Frequency versus $\effective{\scalarW}$ and $\effective{\scalarW^{\dagger}}$. Legend of composition 1 (resp.~2): Re$\,\effective{\scalarW}$  in solid blue (purple) curves; Im$\,\effective{\scalarW}$   in dashed green (pink) curves; Re$\,\effective{\scalarW^\dagger}$  in triangle  (diamond) marks; Im$\,\effective{\scalarW^\dagger}$  in circle   (square) marks. Panel (c) is for composition 1 with $\period^{(b)}=0,\period^{(a)}=\period^{(c)}$, where $\period^{(a)},\period^{(b)}$ and $\period^{(c)}$ denote the thickness of the first, middle and last layer of the unit cell, respectively. }
\label{fig:effectiveL} 
\end{figure}\begin{figure}[t!]
\centering 
\captionsetup[subfloat]{position=top} 
\subfloat[]{\label{fig:Wvsl}		
\includegraphics[width=0.4\textwidth]{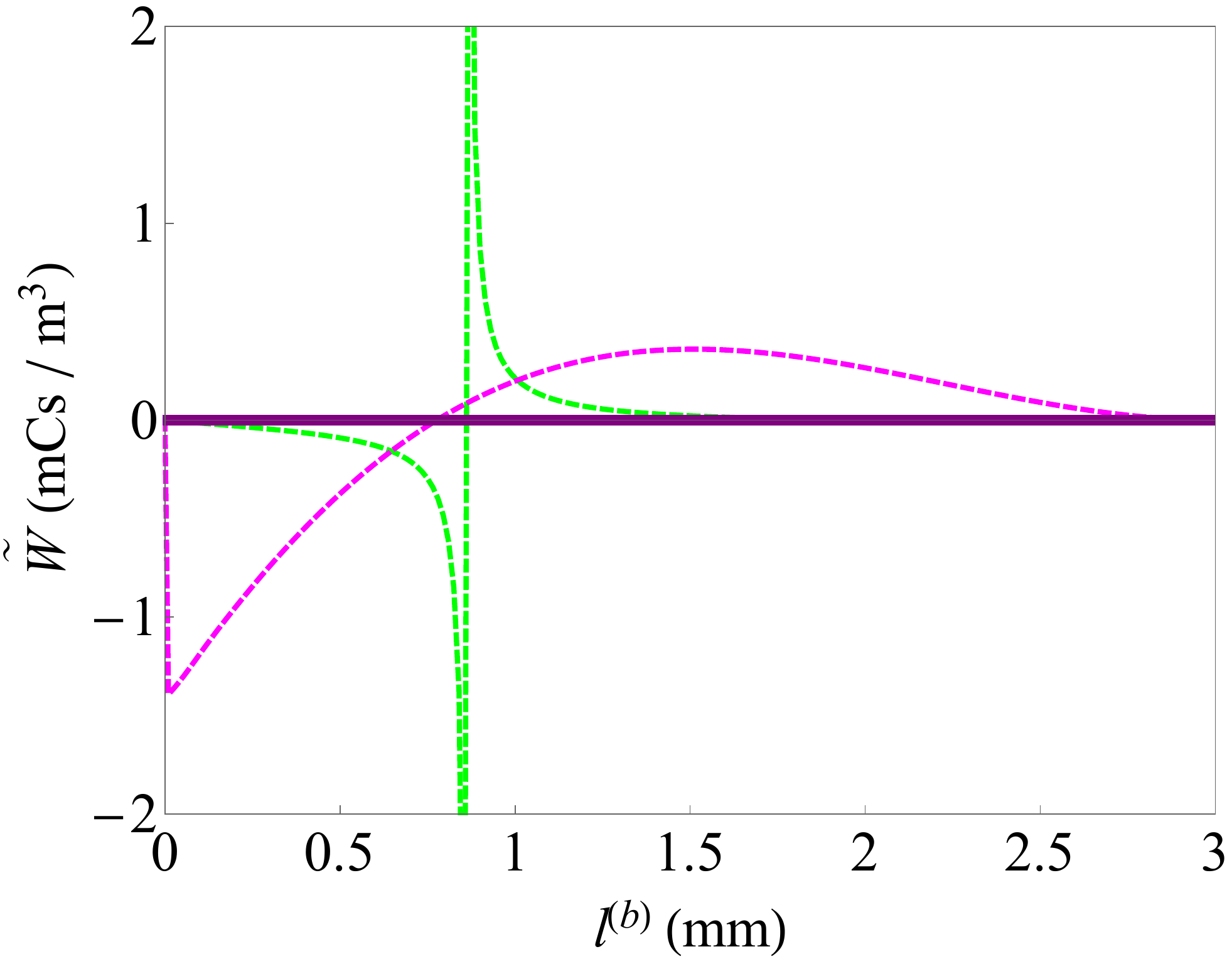}}
\subfloat[]{\label{fig:momentumratio}		
\includegraphics[width=0.4\textwidth]{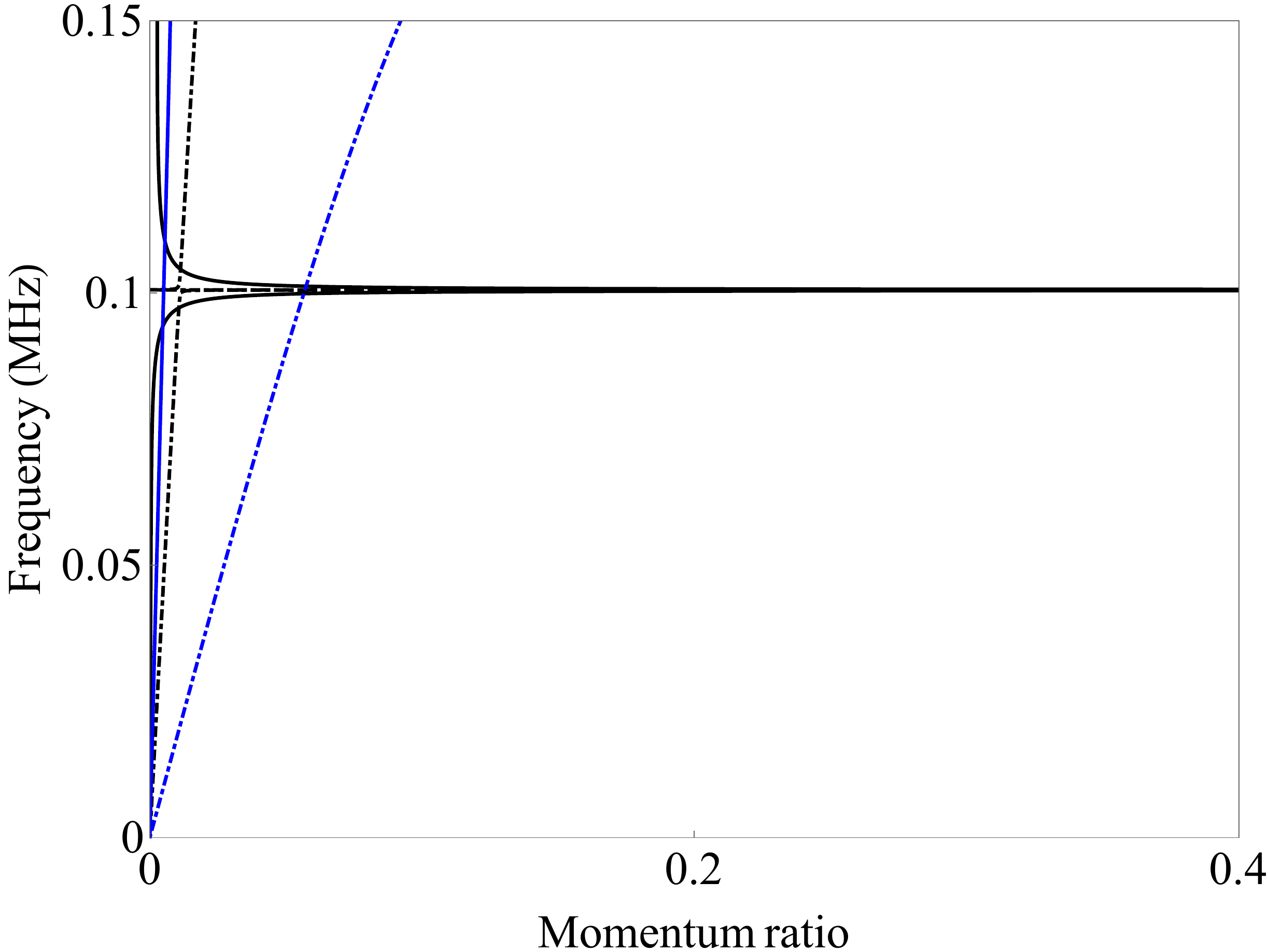}}
\caption{\label{fig:wstudy}
(a) $\effective{\scalarW}$ versus $\period^{(b)}$ at 0.1$\,$MHz and $\xi\period=0$. (b) Frequency versus $\frac{\left|\adjoint{\effective{\scalarW}}\ensemble{\grad\potential}\right|}{\left|\effective{\rho}\ensemble{\dt u}\right|}$ (solid curves) and $\frac{\left|\effective{\scalarS}^{\dagger}\ensemble{\grad u}\right|}{\left|\effective{\rho}\ensemble{\dt u}\right|}$ (dotted curves) at $\xi\period=0$. Black and blue curves correspond to compositions 1 (with $\period^{(b)}=0.87\,$mm) and 2, respectively.}
\end{figure}

Finally, the equivalents of Eqs.~\eqref{eq:effective relation}-\eqref{Eff-L}
are derived directly by substituting $u\left(x,y\right)$ and $\potential\left(x,y\right)$
into Eq.~\eqref{ConstiRel-1}, ensemble averaging, and identifying
the terms that multiply $\ensemble{u_{,x'}}-\eigens,\ensemble{\potential_{,x'}}$
and $\ensemble{su}$ as the effective properties according to Eq.~\eqref{Eff-constR-final}.
The price for using a single Green function in the absence of charge
is one degree of freedom in calculating $\effective{\lmat}$, which
we eliminate by enforcing $\tp{\tilde{\scalarpiezo}}=\tilde{\scalarpiezo}$.
Together with this choice, the resultant equations determine the
components of $\effective{\lmat}$. The explicit expressions are
provided in Appendix D, and imply that \begin{eqnarray}
\begin{aligned}
\effective{\scalarS}\left(\xi\right)&=\adjoint{\effective{\scalarS}}\left(-\xi\right)=-\mathrm{conj}\,\adjoint{\effective{\scalarS}}\left(\xi\right)\\
\effective{\scalarW}\left(\xi\right)&=\adjoint{\effective{\scalarW}}\left(-\xi\right)=-\mathrm{conj}\, \adjoint{\effective{\scalarW}}\left(\xi\right),
\label{adj}
\end{aligned}
\end{eqnarray}where $\mathrm{conj}$ denotes complex conjugate, and $\xi$ is the
Fourier transform variable, \emph{i.e.}, 
\begin{equation}
\tilde{\lmat}\left(\xi\right)=\int_{\cell}\tilde{\lmat}\left(x\right)e^{i\xi x}\mathrm{d}x.\label{eq:fourier transform}
\end{equation}
These relations also imply that Re$\,\effective{\scalarW}$ is an
odd function of $\xi$, while Im$\,\effective{\scalarW}$ is an even
function, similarly to the functional form of $\effective{\scalarS}$
\cite{Sieck2017prb}. Eq.~\eqref{adj} consolidates the notation used
by Willis \cite{Willis2009,Willis2011PRSA,WILLIS2012MOM,WILLIS2012MRC},
who interprets $\effective{\scalarS}$ and $\adjoint{\effective{\scalarS}}$
as formal adjoints with respect to the spatial variable, and other
notations in the literature \cite{Norrisrspa2011PRSA,Srivastava2012prsa,Sieck2017prb},
which use $-\mathrm{conj}\,\effective{\scalarS}$ instead of $\adjoint{\effective{\scalarS}}$. 

The numerical results in Fig.~\ref{fig:effectiveL} display the Fourier
transform $\effective{\scalarW}\left(\xi\right)$ and its adjoint
versus frequency $\frac{\omega}{2\pi}$, where $s=-i\omega$, across
the first pass band. The calculations are done for the microstructure
$\period^{\left(a\right)}=1\,$mm (thickness of the first layer in
the periodic cell) $,\period^{\left(b\right)}=1.4\,$mm (middle layer),
and $\period^{\left(c\right)}=0.6\,$mm (last layer). Panel \ref{fig:xi0}
shows the long-wavelength limit $\xi\period=0$, which contains the
homogenization limit \cite{Srivastava2014WM,srivastava2015elastic},
where in Panel \ref{fig:xi2} we examine $\xi\period=2$. For brevity,
we omit the plots for the remaining effective properties, noting they
are frequency-dependent, real at $\xi\period=0$ (except $\effective{\scalarS}$
and $\effective{\scalarS}^{\dagger}$, which are pure imaginary),
and generally complex for $\xi\period>0$.

Evidently, the contrast between the piezoelectric coefficients of
the layers greatly affects the magnitude of $\effective{\scalarW}$.
For example, the maximal value of Im$\,\effective{\scalarW}$ at $\xi\period=0$
for composition 1 is 21$\:\mu$Cs$\mathrm{m}^{-3}$, while for composition
2 it is over 900$\:\mu$Cs$\mathrm{m}^{-3}$. 

Both compositions share the following notable features. Firstly, the
electro-momentum coupling $\effective{\scalarW}$ vanishes in the
quasi-static limit $\omega\rightarrow0$, as it should. In the long-wavelength
limit and $\omega>0$, it attains non-zero pure imaginary values,
as Eq.~\eqref{adj} implies. Above the long-wavelength limit, the
numerical results confirm that the electro-momentum is complex such
that Im$\effective{\,\scalarW}=$Im$\,\adjoint{\effective{\scalarW}}$
and Re$\effective{\,\scalarW}=-$Re$\,\adjoint{\effective{\scalarW}}$.
Our observations conform with the insights of Sieck et al.~\cite{Sieck2017prb}
on the microstructure-induced Willis coupling: the imaginary part
originates from broken inversion symmetry, hence appears even in the
long-wavelength limit, while mesoscale effects of multiple scattering
create its real part, hence appear at $\xi\period>0$. This is further
demonstrated in Panel ~\ref{fig:xi2sym}, where we evaluate $\effective{\scalarW}$
and $\adjoint{\effective{\scalarW}}$ at $\xi\period=2$ in a system
that is symmetric under inversion by setting $\period^{\left(b\right)}=0$
in composition 1. Indeed, the coupling is pure real in this setting,
and satisfies satisfies $\effective{\scalarW}\left(\xi\right)=-\adjoint{\effective{\scalarW}}\left(\xi\right)$.

Fig.~\ref{fig:Wvsl} directly evaluates the dependency of $\effective{\scalarW}$
on the microstructure, by plotting it against $\period^{(b)}$ while
setting $\period^{(a)}=\period^{(c)}$, at the representative frequency
0.1$\,$MHz. The couplings vanish for $\period^{(b)}=0$ and 3$\,$mm,
as it should for microstructures with inversion symmetry. Notably,
the change from $\period^{(b)}=0$ to $\period^{(b)}=0^{+}$ is discontinuous
for composition 2, as it reflects a shift from an elastic composition
to a piezoelectric one. Interestingly, the coupling of composition
2 vanishes also at $\period^{(b)}=0.79\,$mm, where the coupling of
composition 1 exhibits singularity in the vicinity of $\period^{(b)}=0.87\,$mm.
These values are functions of the frequency; for example, at 0.05$\,$MHz
we calculated lower values, namely, $\period^{(b)}=0.77\,$mm and
$\period^{(b)}=0.84\,$mm, respectively.

Lastly, in Fig.~\ref{fig:momentumratio} we evaluate the frequency
versus the momentum ratios between $\left|\adjoint{\effective{\scalarW}}\ensemble{\grad\potential}\right|,\left|\effective{\scalarS}^{\dagger}\ensemble{\grad u}\right|$
and $\left|\effective{\rho}\ensemble{\dt u}\right|$ at $\xi\period=0$,
for composition 1 with $\period^{(b)}=0.87\,$mm (black curves), and
composition 2 (blue curves) with the microstructure studied in Fig.~\ref{fig:effectiveL}.
Composition 1 exhibits a singularity at 1$\,$MHz, conforming with
the singularity observed in Fig.~\ref{fig:Wvsl}. In the vicinity
of 1$\,$MHz, the growth of $\left|\adjoint{\effective{\scalarW}}\ensemble{\grad\potential}\right|$
is much faster than $\left|\effective{\scalarS}^{\dagger}\ensemble{\grad u}\right|$.
Composition 2 demonstrates lower ratios, as expected from a metamaterial
with an arbitrarily chosen microstructure. 

\section{\label{sec:Summary}Summary and discussion}

We have developed an exact source-driven homogenization scheme for
responsive metamaterials, based on the ensemble averaging approach
of Willis and applied it in a numerical example considering periodic
repetitions of commercially available piezoelectric layers. The scheme
reveals that the effective non-mechanical and momentum-velocity fields
can be coupled by properly designing the microstructure and composition
of the medium. We conjecture that the corresponding couplings are
necessary for obtaining an effective description that respect fundamental
principles such as causality, similarly to the need for Willis coupling
in elastodynamics\textit{\emph{ \cite{Sieck2017prb} and bianisotropic
coupling in electromagnetics }}\cite{Alu2011-PhysRevB,alu2011prb}\textit{\emph{.}}
The new couplings reflect energy conversion between electrical and
mechanical energy in a distinct way than it occurs at the microscale.
Therefore, they capture a new mechanism that can be employed for rectifying
mechanical waves by modulation of external stimuli. We expect that
future studies will show that extraordinary wave response such as
asymmetric reflections and unidirectional transmission are modeled
by the new couplings \cite{pernas2019b}, analogously to Willis coupling
$\ $\cite{Muhlestein2017nc,Merkel2018prb,Melnikov2019nc,Liu2019prx}.
These kind of results are to be guided by theoretical analyses based
upon proper homogenization schemes \cite{quan2018prl}. The present
work opens the route for such studies on the properties of the electro-momentum
coupling and its experimental realizations for metamaterial design. 

\section*{acknowledgements}

We thank anonymous reviewers for their constructive comments that
helped us improve this paper. We acknowledge the support of the Israel
Science Foundation (Grant 1912/15), United States-Israel Binational
Science Foundation (Grant 2014358), and MOST (Grant 880011).  
 Science and Technology (Grant no.~880011).


\section*{Appendix A. Implications on other stimuli-responsive
media }\label{Appendix A}

The structure we developed and its resultant effective description
immediately apply for piezomagnetic media, by observing the following
mathematical connections. The fields $\mathbf{D}$ and $\mathbf{E}$
are equivalent to the magnetic induction and magnetic field (usually
denoted by $\mathbf{B}$ and $\mathbf{H}$, respectively), since they
both satisfy identical differential equations. The latter two fields
are constitutively related by the second-order permeability tensor
(usually denoted by $\mu$), and their coupling with the stress and
strain is captured by the piezomagnetic third-order tensor \cite{milton2002theory}.
Thereby, homogenization of piezomagnetic composites fits exactly into
our scheme, which predicts new macroscopic second-order tensors that
couple the magnetic induction and the velocity, and the momentum and
the magnetic field. 

Implications of our scheme to thermoelasticity are less immediate
and require a separate treatment. However, at the very basic level,
we can draw analogies between $\mathbf{D}$ and the increase of entropy
per unit volume with respect to a reference state (denoted $\vartheta$)
, and between $\mathbf{E}$ and the change in temperature with respect
to some base temperature (denoted $\theta$). The fields $\vartheta$
and $\theta$ are scalar fields that are constitutively coupled through
the constant of specific heat. The microscopic cross-coupling between
$\vartheta$ and $\theta$ to the stress and strain is captured by
the thermal expansion second-order tensor, see Eq.$\ $(2.24) in \cite{milton2002theory}.
We now proceed to the differential equations that govern thermoelasticity.
While $\mathbf{D}$ and $\mathbf{B}$ are subjected to the same differential
equation, there is no such spatial constraint on $\vartheta$; in
some approximated formulations of thermoelasticity, the equations
of heat conduction and energy balance can be combined to form with
the equation of motion two coupled field equations for the temperature
and displacement, see Eqs.$\ $(2.10-23) or Eqs.$\ $(9.6-7)-(9.6-8)
in \cite{hetnarski2008thermal}. We conjecture that homogenizing this
system will expose new macroscopic couplings between $\vartheta$
and the velocity, and between the momentum and $\theta$. 

\section*{\label{Appendix B}Appendix B. Detailed derivation of $\protect\effective{\protect\lmat}$
in index and tensor notation} 
\global\long\def\theequation{B.\arabic{equation}}%
\setcounter{equation}{0}
Eq.~\eqref{Governing-Eqs-Willis-form} in tensor notation is

\begin{equation}
\left(\begin{array}{c}
\div{\Stress}-s^{2}\rho\disp\\
\div{\Edisplacement}
\end{array}\right)+\left(\begin{array}{c}
\force\\
-\charge
\end{array}\right)=\left(\begin{array}{c}
\mathbf{0}\\
0
\end{array}\right).\label{eq:Gov-u}
\end{equation}
The left product of Eq.~\eqref{eq:Gov-u} with $\tp{\green}\text{\ensuremath{\left(\position,\position'\right)}}$
is

\begin{equation}
\left(\begin{array}{c}
\tp{\greentensor_{11}}\cdot\left(\div{\Stress}\right)+\greentensor_{21}\div{\Edisplacement}\\
\greentensor_{12}\cdot\left(\div{\Stress}\right)+\green_{22}\div{\Edisplacement}
\end{array}\right)-\left(\begin{array}{c}
s^{2}\rho\tp{\greentensor_{11}}\cdot\disp\\
s^{2}\rho\greentensor_{12}\cdot\disp
\end{array}\right)+\left(\begin{array}{c}
\tp{\greentensor_{11}}\cdot\force-\greentensor_{21}\charge\\
\greentensor_{12}\cdot\force-\Gindex_{22}\charge
\end{array}\right)=\left(\begin{array}{c}
\mathbf{0}\\
0
\end{array}\right),\label{eq:WGt-tensor}
\end{equation}
or, in index notation,

\begin{equation}
\left(\begin{array}{c}
\tp{\Gindex_{11pj}}\stressindex_{jk,k}+\Gindex_{21p}\Edisplaindex_{k,k}\\
\Gindex_{12j}\stressindex_{jk,k}+\Gindex_{22}\Edisplaindex_{k,k}
\end{array}\right)-\left(\begin{array}{c}
s^{2}\rho\tp{\Gindex_{11pj}}\dispIndex_{j}\\
s^{2}\rho\Gindex_{12j}\dispIndex_{j}
\end{array}\right)+\left(\begin{array}{c}
\tp{\Gindex_{11pj}}\forceindex_{j}-\Gindex_{21p}\thinspace\charge\\
\Gindex_{12j}\forceindex_{j}-\Gindex_{22}\thinspace\charge
\end{array}\right)=\left(\begin{array}{c}
0_{p}\\
0
\end{array}\right).\label{eq:Gtranspose-w}
\end{equation}
The quantities $\green_{11}$, $\green_{12}$ and $\green_{21}$,
and $\green_{22}$ are second order tensor-, vector-, and scalar-valued
functions, respectively, and hence the symbolic $2\times2$ matrix
$\green$ is representable by a $4\times4$ matrix. ($\green_{11}$,
$\green_{12}$ and $\green_{21}$, and $\green_{22}$ are represented
by $3\times3,3\times1,1\times3$ and $1\times1$ blocks, respectively.)
The equation that defines $\green\left(\position,\position'\right)$
is 
\begin{equation}
\dmat^{\mathsf{T}}\lmat\gradt\green=-\delta\mathsf{I},\label{eq:green equation-1}
\end{equation}
where, in tensor notation, the elements of $\delta\mathsf{I}$ are
\[
\delta\mathsf{I}=\left(\begin{array}{cc}
\dirac\left(\position-\position'\right)\identity & 0\\
0 & \dirac\left(\position-\position'\right)
\end{array}\right),
\]
and the elements of $\dmat^{\mathsf{T}}\lmat\gradt\green$ are

\begin{eqnarray}
 & \left(\begin{array}{cc}
\div{\left(\elas:\grad\greentensor_{11}+\tp{\Piezoelectricmodule}\cdot\grad\greentensor_{21}\right)-s^{2}\rho\greentensor_{11}} & \div{\left(\elas:\grad\greentensor_{12}+\tp{\Piezoelectricmodule}\cdot\grad\Gindex_{22}\right)}-s^{2}\rho\greentensor_{12}\\
\div{\left(\Piezoelectricmodule:\grad\greentensor_{11}-\Permittivity\cdot\grad\greentensor_{21}\right)} & \div{\left(\Piezoelectricmodule:\grad\greentensor_{12}-\Permittivity\cdot\grad\Gindex_{22}\right)}
\end{array}\right)\label{eq:dlbg1}
\end{eqnarray}
or, in index notation,

\begin{eqnarray}
 & \left(\begin{array}{cc}
\left\{ \elasindex_{ijkl}\Gindex_{11kp,l}+\tp{\Piezoelectricindex_{ijm}}\Gindex_{21p,m}\right\} {}_{,j}-s^{2}\rho\Gindex_{11ip} & \left\{ \elasindex_{ijkl}\Gindex_{12k,l}+\tp{\Piezoelectricindex_{ijk}}\Gindex_{22,k}\right\} {}_{,j}-s^{2}\rho\Gindex_{12i}\\
\left\{ \Piezoelectricindex_{jnm}\Gindex_{11np,m}-\Permittivityindex_{jm}\Gindex_{21p,m}\right\} {}_{,j} & \left\{ \Piezoelectricindex_{ijk}\Gindex_{12j,k}-\Permittivityindex_{ij}\Gindex_{22,j}\right\} {}_{,i}
\end{array}\right).\label{eq:eq:dlbg2}
\end{eqnarray}
We emphasize that the convention contraction is different when $\green$
is involved. For example, $\elas:\grad\greentensor_{11}$ is $\elasindex_{ijkl}\Gindex_{11kp,l}$
and $\tp{\Piezoelectricmodule}\cdot\grad\greentensor_{21}$ is $\tp{\Piezoelectricindex_{ijm}}\Gindex_{21p,m}$.
The standard convention (contraction with the first two indices of
the right tensor) can be recovered observing that 
\[
\elasindex_{ijkl}\Gindex_{11kp,l}=\tp{\Gindex}_{11pk,l}\elasindex_{klij},
\]
and thus $\elas:\grad\greentensor_{11}=\tp{\grad\greentensor_{11}}:\elas.$
Similarly, 
\[
\tp{\Piezoelectricindex_{ijm}}\Gindex_{21p,m}=\Gindex_{21p,m}\Piezoelectricindex_{mij},
\]
and thus $\tp{\Piezoelectricmodule}\cdot\grad\greentensor_{21}=\grad\greentensor_{21}\cdot\Piezoelectricmodule$.
Using such identities, the standard convention for divergence operation
applies, \emph{i.e}., acting on the last free index.

Right multiplying the transpose of Eq.~\eqref{eq:green equation-1}
with $\genpotential\left(\position\right)$ provides 
\begin{eqnarray}
 & \left(\begin{array}{c}
\left\{ \tp{\Gindex_{11pk,l}}\elasindex_{klij}+\Gindex_{21p,m}\Piezoelectricindex_{mij}\right\} {}_{,j}\dispIndex_{i}+\left\{ \tp{\Gindex_{11pn,m}}\tp{\Piezoelectricindex_{nmj}}-\Gindex_{21p,m}\Permittivityindex_{mj}\right\} {}_{,j}\potential\\
\left\{ \elasindex_{ijkl}\Gindex_{12k,l}+\tp{\Piezoelectricindex_{ijk}}\Gindex_{22,k}\right\} {}_{,j}\dispIndex_{i}+\left\{ \Piezoelectricindex_{ijk}\Gindex_{12j,k}-\Permittivityindex_{ij}\Gindex_{22,j}\right\} {}_{,i}\thinspace\potential
\end{array}\right)\nonumber \\
 & -\left(\begin{array}{c}
\rho s^{2}\tp{\Gindex_{11pi}}\dispIndex_{i}\\
\rho s^{2}\Gindex_{12i}\dispIndex_{i}
\end{array}\right)=-\left(\begin{array}{c}
\dirac_{pi}\dispIndex_{i}(\position)\dirac\left(\position-\position'\right)\\
\potential(\position)\dirac\left(\position-\position'\right)
\end{array}\right).\label{eq:w-app-2}
\end{eqnarray}
Subtracting Eq.~\eqref{eq:w-app-2} from \eqref{eq:Gtranspose-w}
yields 
\begin{eqnarray}
\left(\begin{array}{c}
\tp{\Gindex_{11pj}}\forceindex_{j}-\Gindex_{21p}\thinspace\charge\\
\Gindex_{12j}\forceindex_{j}-\Gindex_{22}\thinspace q
\end{array}\right)+\left(\begin{array}{c}
\tp{\Gindex_{11pj}}\stressindex_{jk,k}+\Gindex_{21p}\Edisplaindex_{k,k}\\
\Gindex_{12j}\stressindex_{jk,k}+\Gindex_{22}\Edisplaindex_{k,k}
\end{array}\right)-\label{eq:difference}\\
\left(\begin{array}{c}
\left\{ \tp{\Gindex_{11pk,l}}\elasindex_{klij}+\Gindex_{21p,m}\Piezoelectricindex_{mij}\right\} {}_{,j}\dispIndex_{i}+\left\{ \tp{\Gindex_{11pn,m}}\tp{\Piezoelectricindex_{nmj}}-\Gindex_{21p,m}\Permittivityindex_{mj}\right\} {}_{,j}\potential\\
\left\{ \elasindex_{ijkl}\Gindex_{12k,l}+\tp{\Piezoelectricindex_{ijk}}\Gindex_{22,k}\right\} {}_{,j}\thinspace\dispIndex_{i}+\left(\Piezoelectricindex_{ijk}\Gindex_{12j,k}-\Permittivityindex_{ij}\Gindex_{22,j}\right){}_{,i}\potential
\end{array}\right) & =\left(\begin{array}{c}
\dirac_{pi}\dispIndex_{i}(\position)\dirac\left(\position-\position'\right)\\
\potential(\position)\dirac\left(\position-\position'\right)
\end{array}\right) & .\nonumber 
\end{eqnarray}
Eq.~\eqref{eq:difference} is simplified using the following relations

\begin{eqnarray*}
\tp{\Gindex_{11pj}}\stressindex_{jk,k}+\Gindex_{21p}\Edisplaindex_{k,k} & = & \left\{ \tp{\Gindex_{11pj}}\stressindex_{jk}\right\} {}_{,k}-\tp{\Gindex_{11pj,k}}\stressindex_{jk}+\left\{ \Gindex_{21p}\Edisplaindex_{k}\right\} {}_{,k}-\Gindex_{21p,k}\Edisplaindex_{k}\\
 & = & \left\{ \tp{\Gindex_{11pj}}\stressindex_{jk}+\Gindex_{21p}\Edisplaindex_{k}\right\} {}_{,k}-\tp{\Gindex_{11pj,k}}\left[\elasindex_{jkil}(\dispIndex_{i,l}-\eigenindex_{il})+\tp{\Piezoelectricindex_{jki}}\potential_{,i}\right]\\
 &  & -\Gindex_{21p,k}\left[\Piezoelectricindex_{kil}\left(\dispIndex_{i,l}-\eigenindex_{il}\right)-\Permittivityindex_{ki}\potential_{,i}\right]\\
 & = & \left\{ \tp{\Gindex_{11pj}}\stressindex_{jk}+\Gindex_{21p}\Edisplaindex_{k}\right\} {}_{,k}-\left(\tp{\Gindex_{11pj,k}}\elasindex_{jkil}+\Gindex_{21p,k}\Piezoelectricindex_{kil}\right)\dispIndex_{i,l}\\
 &  & -\left(\tp{\Gindex_{11pj,k}}\tp{\Piezoelectricindex_{jki}}-\Gindex_{21p,k}\Permittivityindex_{ki}\right)\potential_{,i}+\left(\tp{\Gindex_{11pj,k}}\elasindex_{jkil}+\Gindex_{21p,k}\Piezoelectricindex_{kil}\right)\eigenindex_{il};\\
\Gindex_{12j}\stressindex_{jk,k}+\Gindex_{22}\Edisplaindex_{k,k} & = & \left\{ \Gindex_{12j}\stressindex_{jk}\right\} {}_{,k}-\Gindex_{12j,k}\stressindex_{jk}+\left\{ \Gindex_{22}\Edisplaindex_{k}\right\} {}_{,k}-\Gindex_{22,k}\Edisplaindex_{k}\\
 & = & \left\{ \Gindex_{12j}\stressindex_{jk}+\Gindex_{22}\Edisplaindex_{k}\right\} {}_{,k}-\Gindex_{12j,k}\left[\elasindex_{jkil}(\dispIndex_{i,l}-\eigenindex_{il})+\tp{\Piezoelectricindex_{jki}}\potential_{,i}\right]\\
 &  & -\Gindex_{22,k}\left[\Piezoelectricindex_{kil}\left(\dispIndex_{i,l}-\eigenindex_{il}\right)-\Permittivityindex_{ki}\potential_{,i}\right]\\
 & = & \left\{ \Gindex_{12j}\stressindex_{jk}+\Gindex_{22}\Edisplaindex_{k}\right\} {}_{,k}-\left(\elasindex_{iljk}\Gindex_{12j,k}+\tp{\Piezoelectricindex_{ilk}}\Gindex_{22,k}\right)\dispIndex_{i,l}\\
 &  & -\left(\Piezoelectricindex_{ijk}\Gindex_{12j,k}-\Permittivityindex_{ik}\Gindex_{22,k}\right)\potential_{,i}+\left(\Gindex_{12j,k}\elasindex_{jkil}+\Gindex_{22,k}\Piezoelectricindex_{kil}\right)\eigenindex_{il};\\
\left\{ \tp{\Gindex_{11pk,l}}\elasindex_{klij}+\Gindex_{21p,m}\Piezoelectricindex_{mij}\right\} {}_{,j}\dispIndex_{i} & = & \left\{ \left(\tp{\Gindex_{11pk,l}}\elasindex_{klij}+\Gindex_{21p,m}\Piezoelectricindex_{mij}\right)\dispIndex_{i}\right\} {}_{,j}-\left(\tp{\Gindex_{11pk,l}}\elasindex_{klij}+\Gindex_{21p,m}\Piezoelectricindex_{mij}\right)\dispIndex_{i,j};\\
\left\{ \tp{\Gindex_{11pn,m}}\tp{\Piezoelectricindex_{nmj}}-\Gindex_{21p,m}\Permittivityindex_{mj}\right\} {}_{,j}\potential & = & \left\{ \left(\tp{\Gindex_{11pn,m}}\tp{\Piezoelectricindex_{nmj}}-\Gindex_{21p,m}\Permittivityindex_{mj}\right)\potential\right\} {}_{,j}-\left(\tp{\Gindex_{11pn,m}}\tp{\Piezoelectricindex_{nmj}}-\Gindex_{21p,m}\Permittivityindex_{mj}\right)\potential_{,j};\\
\left\{ \elasindex_{ijkl}\Gindex_{12k,l}+\tp{\Piezoelectricindex_{ijk}}\Gindex_{22,k}\right\} {}_{,j}\dispIndex_{i} & = & \left\{ \left(\elasindex_{ijkl}\Gindex_{12k,l}+\tp{\Piezoelectricindex_{ijk}}\Gindex_{22,k}\right)\dispIndex_{i}\right\} {}_{,j}-\left(\elasindex_{ijkl}\Gindex_{12k,l}+\tp{\Piezoelectricindex_{ijk}}\Gindex_{22,k}\right)\dispIndex_{i,j};\\
\left\{ \Piezoelectricindex_{ijk}\Gindex_{12j,k}-\Permittivityindex_{ij}\Gindex_{22,j}\right\} {}_{,i}\potential & = & \left\{ \left(\Piezoelectricindex_{ijk}\Gindex_{12j,k}-\Permittivityindex_{ij}\Gindex_{22,j}\right)\potential\right\} {}_{,i}-\left(\Piezoelectricindex_{ijk}\Gindex_{12j,k}-\Permittivityindex_{ij}\Gindex_{22,j}\right)\potential_{,i},
\end{eqnarray*}
which, upon integration over the volume $\position\in\vol$ and application
of divergence theorem, then reads

\begin{eqnarray}
\left(\begin{array}{c}
\dispIndex_{p}(\position')\\
\potential(\position')
\end{array}\right) & = & \intvol{\left(\begin{array}{c}
\tp{\Gindex_{11pj}}\forceindex_{j}-\Gindex_{21p}\thinspace\charge\\
\Gindex_{12j}\forceindex_{j}-\Gindex_{22}\thinspace q
\end{array}\right)}+\intsurfaceh{\left(\begin{array}{c}
\left(\tp{\Gindex_{11pj}}\stressindex_{jk}+\Gindex_{21p}\Edisplaindex_{k}\right)\nhindex_{k}\\
\left(\Gindex_{12j}\stressindex_{jk}+\Gindex_{22}\Edisplaindex_{k}\right)\nhindex_{k}
\end{array}\right)}\nonumber \\
 & + & \intvol{\left(\begin{array}{c}
\left(\tp{\Gindex_{11pj,k}}\elasindex_{jkil}+\Gindex_{21p,k}\Piezoelectricindex_{kil}\right)\eigenindex_{il}\\
\left(\Gindex_{12j,k}\elasindex_{jkil}+\Gindex_{22,k}\Piezoelectricindex_{kil}\right)\eigenindex_{il}
\end{array}\right)}\\
 & - & \intsurfacepot{\left(\begin{array}{c}
\left[\left(\tp{\Gindex_{11pk,l}}\elasindex_{klij}+\Gindex_{21p,m}\Piezoelectricindex_{mij}\right)\dispIndex_{i}\right]\nhindex_{j}+\left[\left(\tp{\Gindex_{11pn,m}}\tp{\Piezoelectricindex_{nmj}}-\Gindex_{21p,m}\Permittivityindex_{mj}\right)\potential\right]\nhindex_{j}\\
\left[\left(\elasindex_{ijkl}\Gindex_{12k,l}+\tp{\Piezoelectricindex_{ijk}}\Gindex_{22,k}\right)\dispIndex_{i}\right]\nhindex_{j}+\left[\left(\Piezoelectricindex_{ijk}\Gindex_{12j,k}-\Permittivityindex_{ij}\Gindex_{22,j}\right)\potential\right]\nhindex_{i}
\end{array}\right)},\nonumber 
\end{eqnarray}
or, in tensor notation,

\begin{eqnarray}
\left(\begin{array}{c}
\disp(\position')\\
\potential(\position')
\end{array}\right) & = & \intvol{\left(\begin{array}{c}
\tp{\greentensor_{11}}\cdot\force-\greentensor_{21}\charge\\
\greentensor_{12}\cdot\force-\Gindex_{22}\charge
\end{array}\right)}+\intsurfaceh{\left(\begin{array}{c}
\tp{\greentensor_{11}}\cdot\left(\Stress\cdot\nh\right)+\greentensor_{21}\left(\Edisplacement\cdot\nh\right)\\
\greentensor_{12}\cdot\left(\Stress\cdot\nh\right)+\Gindex_{22}\left(\Edisplacement\cdot\nh\right)
\end{array}\right)}\nonumber \\
 & + & \intvol{\left(\begin{array}{c}
\left(\tp{\grad\greentensor_{11}}:\elas+\grad\greentensor_{21}\cdot\Piezoelectricmodule\right):\eigen\\
\left(\grad\greentensor_{12}:\elas+\grad\Gindex_{22}\cdot\Piezoelectricmodule\right):\eigen
\end{array}\right)}\label{eq:w-tensor-notation}\\
 & - & \intsurfacepot{\left(\begin{array}{c}
\left(\tp{\grad\greentensor_{11}}:\elas+\grad\greentensor_{21}\cdot\Piezoelectricmodule\right):\disp\otimes\nh+\left(\tp{\grad\greentensor_{11}}:\tp{\Piezoelectricmodule}-\grad\greentensor_{21}\cdot\Permittivity\right)\cdot\potential\nh\\
\left(\grad\greentensor_{12}:\elas+\grad\Gindex_{22}\cdot\Piezoelectricmodule\right):\disp\otimes\nh+\left(\grad\greentensor_{12}:\tp{\Piezoelectricmodule}-\grad\Gindex_{22}\cdot\Permittivity\right)\cdot\potential\nh
\end{array}\right)},\nonumber 
\end{eqnarray}
where the tensor product $\otimes$ between vectors $\mathbf{a}$
and $\mathbf{b}$ is defined by the action on a third vector $\mathbf{c}$,
namely \cite{ogden97book}, 
\[
\left(\mathbf{a}\otimes\mathbf{b}\right)\cdot\mathbf{c}=\left(\mathbf{b}\cdot\mathbf{c}\right)\mathbf{a},
\]
implying that $\left(\mathbf{a}\otimes\mathbf{b}\right)_{ij}=a_{i}b_{j}$
and $\left(\mathbf{T}\otimes\mathbf{b}\right)_{ijk}=T_{ij}b_{k}$
for second-order tensors $\mathbf{T}$. Accordingly, 
\begin{align*}
\left(\tp{\greentensor_{11}}\otimes\nh\right):\Stress & =\tp{\greentensor_{11}}\cdot\left(\Stress\cdot\nh\right),\\
\left(\greentensor_{21}\otimes\nh\right)\cdot\Edisplacement & =\greentensor_{21}\left(\Edisplacement\cdot\nh\right),\\
\left(\greentensor_{12}\otimes\nh\right):\Stress & =\greentensor_{12}\cdot\left(\Stress\cdot\nh\right),\\
\left(\Gindex_{22}\otimes\nh\right)\cdot\Edisplacement & =\Gindex_{22}\left(\Edisplacement\cdot\nh\right),
\end{align*}
which, by defining 
\[
\tp{\nmat}=\left(\begin{array}{ccc}
\otimes\nh & 0 & 0\\
0 & \otimes\nh & 0
\end{array}\right),
\]
allows us to write Eq.~\eqref{eq:w-tensor-notation} in the symbolic
matrix form

\begin{eqnarray}
\genpotential(\position') & = & \intvol{\tp{\green}\fmat}+\intsurfaceh{\tp{\green}\tp{\nmat}\lmat\left(\gradt\genpotential-\mmat\right)}\nonumber \\
 & + & \intvol{\tp{\left(\gradt\green\right)}\lmat\mmat}-\intsurfacepot{\tp{\left(\gradt\green\right)}\lmat\nmat\genpotential}.\label{eq:w-appendix-2}
\end{eqnarray}
We analyze next the last integral. First, we formally extend its domain
from $\boundaryw$ to the whole boundary $\bnd$ using the fact that
the integrand vanishes over $\boundaryt$, owing to homogeneous boundary
conditions 
\begin{equation}
\tp{\left(\gradt\green\right)}\lmat\nmat=0\ \mathrm{over}\ \boundaryt\label{eq:gbcappendix}
\end{equation}
that $\green$ satisfies over $\boundaryt$. Next, we can replace
$\genpotential$ with $\ensemble{\genpotential}$, since it is sure
on $\boundaryw$ and the integrand vanishes where it is not. By transforming
the surface integral back to the volume via the divergence theorem
and employing Eq.~\eqref{eq:green equation}, we rewrite the second
line as 

\begin{eqnarray}
 &  & \intvol{\tp{\left(\gradt\green\right)}\lmat\mmat}-\intvol{\left[\tp{\left(\dmat^{\mathsf{T}}\lmat\gradt\green\right)}\ensemble{\genpotential}+\tp{\left(\gradt\green\right)}\lmat\gradt\ensemble{\genpotential}\right]}=\nonumber \\
 &  & -\intvol{\tp{\left(\dmat^{\mathsf{T}}\lmat\gradt\green\right)}\ensemble{\genpotential}}-\intvol{\tp{\left(\gradt\green\right)}\lmat\left(\gradt\ensemble{\genpotential}-\mmat\right)}=\nonumber \\
 &  & \intvol{\tp{\left(\mathsf{\delta}\mathsf{I}\right)}\ensemble{\genpotential}}-\intvol{\tp{\left(\gradt\green\right)}\lmat\left(\gradt\ensemble{\genpotential}-\mmat\right)}=\nonumber \\
 &  & \ensemble{\genpotential}(\position')-\intvol{\tp{\left(\gradt\green\right)}\lmat\left(\gradt\ensemble{\genpotential}-\mmat\right)},\label{eq:A0}
\end{eqnarray}
Now, since $\fmat$ is sure over $\vol$ and $\tp{\nmat}\lmat\left(\gradt\genpotential-\mmat\right)$
is sure over $\boundaryt$ (it is the surface charge and traction),
we can combine the ensemble average of Eqs.~ \eqref{eq:w-appendix-2}
and \eqref{eq:A0} to obtain 
\begin{equation}
\intvol{\ensemble{\tp{\green}}\fmat}+\intsurfaceh{\ensemble{\tp{\green}}\tp{\nmat}\lmat\left(\gradt\genpotential-\mmat\right)}=\intvol{\ensemble{\tp{\left(\gradt\green\right)}\lmat}\left(\gradt\ensemble{\genpotential}-\mmat\right)},\label{eq:A0-2}
\end{equation}
and hence

\begin{equation}
\intvol{\tp{\green}\fmat}+\intsurfaceh{\tp{\green}\tp{\nmat}\lmat\left(\gradt\genpotential-\mmat\right)}=\intvol{\tp{\green}\ensemble{\green}^{-\mathsf{T}}\ensemble{\tp{\left(\gradt\green\right)}\lmat}\left(\gradt\ensemble{\genpotential}-\mmat\right)}.\label{eq:A0-3}
\end{equation}
Finally, the above manipulations allow us to write $\genpotential\left(\position'\right)$
as 
\begin{equation}
\genpotential(\position')=\ensemble{\genpotential}(\position')-\intvol{\tp{\left(\gradt\green\right)}\lmat\left(\gradt\ensemble{\genpotential}-\mmat\right)}+\intvol{\tp{\green}\ensemble{\green}^{-\mathsf{T}}\ensemble{\tp{\left(\gradt\green\right)}\lmat}\left(\gradt\ensemble{\genpotential}-\mmat\right)}.\label{eq:w-appendix-final}
\end{equation}
In essence, Eq.$\ $\eqref{eq:w-appendix-final} is the generalization
of Eq.$\ $(3.14) by Willis \cite{Willis2011PRSA} to piezoelectric
media, and as such, relies upon similar reasoning in its derivation.
The effective operator $\effective{\lmat}$ in our settings is obtained
by substituting Eq.~\eqref{eq:w-appendix-final} into $\hmat\left(\position'\right)=\lmat\left(\gradt\genpotential-\mmat\right)\left(\position'\right)$,
namely,

\begin{eqnarray}
\hmat(\position') & = & \intvol{\dirac\left(\position-\position'\right)\lmat(\position')\left\{ \gradt\ensemble{\genpotential}(\position)-\mmat(\position)\right\} }-\intvol{\lmat(\position')\gradt\left(\position'\right)\tp{\left(\gradt\green\right)}\lmat(\position)\left\{ \gradt\ensemble{\genpotential}(\position)-\mmat\left(\position\right)\right\} }\nonumber \\
 & + & \intvol{\lmat(\position')\gradt\left(\position'\right)\tp{\green}\ensemble{\green}^{-\mathsf{T}}\ensemble{\tp{\left(\gradt\green\right)}\lmat}\left\{ \gradt\ensemble{\genpotential}(\position)-\mmat(\position)\right\} }.\label{eq:hxp}
\end{eqnarray}
Finally, ensemble averaging Eq.~\eqref{eq:hxp} and comparing it
with $\ensemble{\hmat}=\effective{\lmat}\left(\ensemble{\gradt\genpotential}-\mmat\right)$
delivers the following expression for $\effective{\lmat}$

\begin{equation}
\effective{\lmat}=\ensemble{\lmat}-\ensemble{\lmat\gradt\tp{\left(\gradt\green\right)}\lmat}+\ensemble{\lmat\gradt\tp{\green}}\ensemble{\green}^{-\mathsf{T}}\ensemble{\tp{\left(\gradt\green\right)}\lmat}.
\end{equation}

\global\long\def\theequation{C.\arabic{equation}}%
 \setcounter{equation}{0}

\section*{\label{Appendix C}Appendix C. Construction of the Green function
using Floquet solutions }

\begin{figure}[t]
\includegraphics[width=1\textwidth]{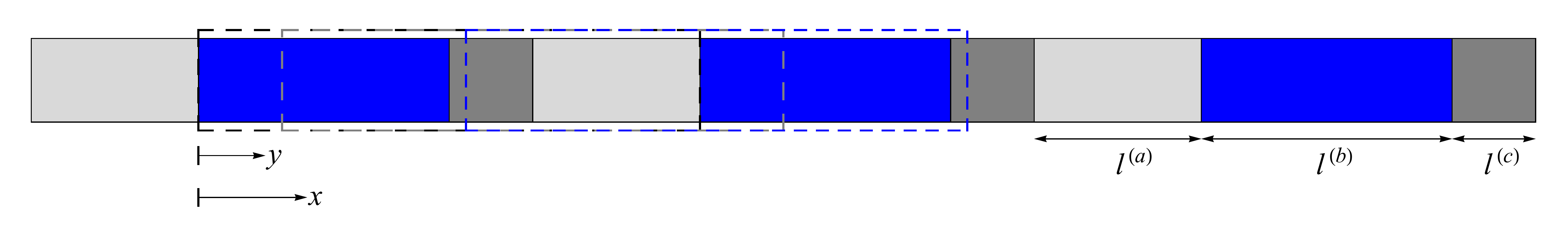}\caption{Part of an infinite medium made of a periodic cell with three layers.
The medium can be analyzed as random, when considering different realizations
generated by periodic cells whose corner varies with $y$---a uniformly
distributed variable; three of which are boxed in dashed lines. }

{\small{}{}\label{fig:realizations}}{\small\par}
\end{figure}

\noindent Recall that in realization $\realization=0$, the governing
equations are combined to obtain in each layer 
\begin{equation}
\left(\zerosub{\scalarelas}+\dfrac{\zerosub{\scalarpiezo}^{2}}{\zerosub{\scalarA}}\right)\left(u_{0,xx}-\eta_{,x}\right)-s^{2}\zerosub{\rho}u_{0}=-\scalarf_{0}.\label{eq:simplified eom-1}
\end{equation}
Since the homogeneous equation derived from Eq.~\eqref{eq:simplified eom-1}
has periodic coefficients, its Green function is constructed using
Bloch-Floquet solutions, namely \cite{Eastham1973j}, 
\begin{eqnarray}
\begin{aligned}\zerosub G\left(x,x^{\prime}\right) & =\left\lbrace \begin{array}{l}
\wronskian\pl u\left(x\right)\mn u(x^{\prime}),\quad x<x^{\prime},\\[10pt]
\wronskian\pl u(x^{\prime})\mn u(x),\ \quad x^{\prime}<x,
\end{array}\right.\label{GreenF}\end{aligned}
\end{eqnarray}
where 
\begin{equation}
\wronskian^{-1}=\left(\zerosub{\scalarelas}+\dfrac{\zerosub{\scalarpiezo}^{2}}{\zerosub{\scalarA}}\right)\left(\pl u_{,x}\mn u-\pl u\mn u_{,x}\right),\ u^{\pm}=\periodic u^{\pm}\left(x\right)e^{\pm i\kb x},\ \periodic u^{\pm}\left(x+l\right)=\periodic u^{\pm}\left(x\right).
\end{equation}
Before we proceed to calculate the periodic parts $\periodic u^{\pm}\left(x\right)$,
it is worth showing that ensemble averaging of Bloch-Floquet functions
reduces to volume averaging over their periodic part \textit{\emph{\cite{Willis2011PRSA,srivastava2015elastic,nassar2015willis}.
To this end, observe that the solution, say $\sol$, to a differential
equation with $l$-periodic coefficients and Bloch wavenumber $\kb$
governing realization $y$ is of the form 
\begin{equation}
\sol\left(x,y,t\right)=\sol_{y}\left(x\right)e^{i\left(k_{B}x-\omega t\right)},\ \sol_{y}\left(x+l\right)=\sol_{y}\left(x\right),\label{eq:floquetform}
\end{equation}
We now recall that $\mathcal{U}_{y}\left(x\right)$ in realization
$y$ is related to realization 0 via (Fig.$\ $\ref{fig:realizations})
\begin{equation}
\sol_{y}\left(x\right)=\sol_{0}\left(x-y\right),\label{eq:shift}
\end{equation}
and therefore ensemble averaging over $\Omega_{\mathrm{p}}$ amount
to 
\begin{equation}
\ensemble{\sol}=\frac{1}{\period}\ensembleint{\sol\left(x,y,t\right)}=\frac{1}{\period}e^{i\left(k_{B}x-\omega t\right)}\ensembleint{\sol\left(x-\realization\right)},\label{eq:ensembleexample}
\end{equation}
}}

\noindent \textit{\emph{thereby equals to volume averaging of the
period part. It is straightforward to extend these arguments to the
three-dimensional case.}}

\noindent We return to calculate $\periodic u^{\pm}\left(x\right)$
by employing a standard transfer matrix approach \cite{Dunkin1965,Shmuel2016sms},
as follows. The phase-wise solution of the Eq.~\eqref{eq:simplified eom-1}
is 
\begin{eqnarray}
u(\positionscalar)=\left\lbrace \begin{array}{l}
u^{\left(a\right)}=\constanta^{(a)}\cos k^{(a)}\positionscalar+\constantb^{(a)}\sin k^{(a)}\positionscalar,\;\;\;-l^{(a)}<\positionscalar<0,\\[3pt]
u^{\left(b\right)}=\constanta^{(b)}\cos k^{(b)}\positionscalar+\constantb^{(b)}\sin k^{(b)}\positionscalar,\;\;\;0<\positionscalar<l^{(b)},\\[3pt]
u^{\left(c\right)}=\constanta^{(c)}\cos k^{(c)}\positionscalar+\constantb^{(c)}\sin k^{(c)}\positionscalar,\;\;\;l^{(b)}<\positionscalar<l^{(b)}+l^{(c)},
\end{array}\right.\label{u-app}
\end{eqnarray}
where $\phase{\constanta}$ and $\phase{\constantb}$ are integration
constants, 
\[
\phase k=s\sqrt{\dfrac{\phase{\rho}}{\phase{\equivc}}},\quad\phase{\equivc}=\phase{\elascalar}+\dfrac{\left\{ \phase{\Piezoelectricscalar}\right\} ^{2}}{\phase{\Permittivityscalar}},\ n=a,b,c,
\]
and to abbreviate notation, we momentarily suppress the subscript
$0$. The integration constants $\phase{\constanta}$ and $\phase{\constantb}$
are determined from the continuity and Bloch-Floquet conditions on
$u(\positionscalar)$ and the stress $\stressindex(\positionscalar)$,
which are compactly written using the state vectors $\phase{\statev}\left(x\right)$
and transfer matrices $\phase{\transfer}$ 
\begin{equation}
\phase{\statev}\left(x\right)=\left(\begin{array}{c}
\phase u\\
\phase{\stressindex}
\end{array}\right),\quad\phase{\transfer}=\left(\begin{array}{cc}
\cos\phase k\phase l & \dfrac{\sin\phase k\phase l}{\phase{\equivc}\phase k}\\
-\phase{\equivc}\phase k\sin\phase k\phase l & \cos\phase k\phase l
\end{array}\right).\label{eq:statetransfer}
\end{equation}
Thus, the state vectors at the ends of each layer are related via
\begin{equation}
\phase{\statev}\left(\phase x+\phase{\period}\right)=\phase{\transfer}\phase{\statev}\left(\phase x\right),\quad\phase x=-\period^{\left(a\right)},0,\period^{\left(b\right)}.\label{eq:T}
\end{equation}
where continuity conditions at the interfaces are 
\begin{equation}
\phase{\statev}\left(\phase x+\phase{\period}\right)=\statev^{\left(n+1\right)}\left(\phase x+\phase{\period}\right),\quad\phase x=0,\period^{\left(b\right)},\label{eq:cont}
\end{equation}
which are combined to obtain 
\begin{eqnarray}
\statev^{\left(c\right)}\left(\period^{\left(c\right)}\right)=\transfer^{\cell}\statev^{\left(a\right)}\left(-\period^{\left(a\right)}\right), & \transfer^{\cell}=\transfer^{(c)}\transfer^{(b)}\transfer^{(a)}.\label{uc-matrix}
\end{eqnarray}
The latter quantities are also related via the (Bloch) Floquet quasi-periodicity
condition 
\begin{equation}
\statev^{\left(c\right)}\left(\period^{\left(c\right)}\right)=e^{i\kb l}\statev^{\left(a\right)}\left(-\period^{\left(a\right)}\right),\label{eq:blochcondition}
\end{equation}
Eqs.~\eqref{uc-matrix}-\eqref{eq:blochcondition} deliver together
the eigenvalue problem 
\begin{equation}
\left(\transfer^{\cell}-e^{i\Blochwn l}\bI\right)\statev=\mathsf{0}.\label{eq:dispersion}
\end{equation}
The standard condition for non-trivial solutions provides the dispersion
relation 
\begin{equation}
\cos\kb l=\frac{\mathrm{tr}\transfer^{\cell}}{2},\label{Exact-dispersion-relation}
\end{equation}
which, upon substitution back into the foregoing equations, provides
the eigenmodes as functions of $\kb$, and specifically $u^{\pm}=\periodic u^{\pm}\left(x\right)e^{\pm i\kb x}$.

\noindent In the calculations to follow is more convenient to use
the Fourier expansion of of $\pl{\periodic u}$ and $\mn{\periodic u}$.
Accordingly, in terms of the Fourier coefficients 
\begin{equation}
\begin{aligned}a_{m}(\pm\Blochwn) & =\dfrac{1}{l}\int_{-l^{(a)}}^{l^{(b)}+l^{(c)}}u^{\pm}(x)e^{\mp i\Blochwn\positionscalar}e^{\frac{2i\pi m\positionscalar}{l}}\mathrm{d}\positionscalar,\\[5pt]\end{aligned}
\label{eq:am}
\end{equation}

\noindent the ensemble average of $\scalarG$ 
\begin{equation}
\ensemble{\scalarG}\left(x,x'\right)=\frac{1}{\period}\ensembleint{\scalarG_{y}\left(x,x'\right)},\quad\scalarG_{y}\left(x,x'\right)=\scalarG_{0}\left(x-\realization,x'-\realization\right),\label{eq:ensg}
\end{equation}
namely, 
\begin{equation}
\ensemble{\scalarG}\left(\positionscalar,\positionscalar^{\prime}\right)=\coeffG\,e^{-i\Blochwn\lvert\positionscalar-\positionscalar^{\prime}\lvert}\sum_{m=-\infty}^{m=\infty}a_{m}(\Blochwn)a_{-m}(-\Blochwn)\,e^{\frac{2i\pi m\lvert\positionscalar-\positionscalar^{\prime}\lvert}{l}},
\end{equation}
confirming it depends on $\positionscalar$ and $x'$ solely via $x-x'$.
Its Fourier transform 
\begin{equation}
\ensemble{\scalarG}\left(\xi\right)=\int_{\cell}\ensemble{\scalarG}\left(x\right)e^{i\xi x}\label{eq:fouriertransform}
\end{equation}
reads 
\begin{equation}
\ensemble{\scalarG}(\xi)=2\coeffG\,\sum_{m=-\infty}^{m=\infty}a_{m}(\Blochwn)a_{-m}(-\Blochwn)\dfrac{i\left(\Blochwn-\dfrac{2\pi m}{l}\right)}{\xi^{2}-\left(\Blochwn-\dfrac{2\pi m}{l}\right)^{2}}.
\end{equation}

We proceed to calculate the rest of the terms in the Fourier transform
of $\effective{\lmat}.$ To this end, we define the Fourier coefficients
\begin{align}
\rho_{m}(\pm\Blochwn) & =\dfrac{1}{l}\int_{-l^{(a)}}^{l^{(b)}+l^{(c)}}\rho_{0}(\positionscalar)e^{\mp i\Blochwn\positionscalar}\plmn u(\positionscalar)e^{\frac{2i\pi m\positionscalar}{l}}\mathrm{d}\positionscalar,\\
\coefficients_{m}(\pm\Blochwn) & =\dfrac{1}{l}\int_{-l^{(a)}}^{l^{(b)}+l^{(c)}}\coefficients_{0}(\positionscalar)e^{\mp i\Blochwn\positionscalar}\plmn u_{,x}(\positionscalar)e^{\frac{2i\pi m\positionscalar}{l}}\mathrm{d}\positionscalar,\quad\coefficients=\elascalar,\check{\elascalar},\dfrac{\Piezoelectricscalar}{\Permittivityscalar},\dfrac{\Piezoelectricscalar^{2}}{\Permittivityscalar},
\end{align}
and obtain

\begin{eqnarray}
\ensemble{\coefficients(\positionscalar)\scalarG{}_{,\positionscalar}}(\xi) & = & \coeffG\sum_{m=-\infty}^{m=\infty}\left[\dfrac{\coefficients_{m}(\Blochwn)a_{-m}(-\Blochwn)}{\adjointerm_{m}^{+}}+\dfrac{\coefficients_{-m}(-\Blochwn)a_{m}(\Blochwn)}{\adjointerm_{m}^{-}}\right],\label{eq:zg}\\
\\
\ensemble{\coefficients_{\mathrm{\left(1\right)}}(\positionscalar)\scalarG{}_{,\positionscalar}\coefficients_{\mathrm{\mathrm{\left(2\right)}}}(\positionscalar^{\prime})}(\xi) & = & \coeffG\sum_{m=-\infty}^{m=\infty}\left[\dfrac{\coefficients_{\mathrm{\mathrm{\left(1\right)}}m}(\Blochwn)\coefficients_{\mathrm{\left(2\right)}-m}(-\Blochwn)}{\adjointerm_{m}^{+}}+\dfrac{\coefficients_{\mathrm{\mathrm{\left(1\right)}-m}}(-\Blochwn)\coefficients_{\mathrm{\left(2\right)}m}(\Blochwn)}{\adjointerm_{m}^{-}}\right],\\
\ensemble{\coefficients(\positionscalar)\scalarG{}_{,\positionscalar\positionscalar^{\prime}}\coefficients(\positionscalar^{\prime})}(\xi) & = & \coeffG\sum_{m=-\infty}^{m=\infty}\left[\dfrac{\coefficients_{m}(\Blochwn)\coefficients_{-m}(-\Blochwn)}{\adjointerm_{m}^{+}}+\dfrac{\coefficients_{\mathrm{-m}}(-\Blochwn)\coefficients_{m}(\Blochwn)}{\adjointerm_{m}^{-}}\right]+\ensemble{\coefficients},\\
\ensemble{\coefficients_{\mathrm{\left(1\right)}}(\positionscalar)\scalarG{}_{,\positionscalar\positionscalar^{\prime}}\coefficients_{\mathrm{\left(2\right)}}(\positionscalar^{\prime})}(\xi) & = & \coeffG\sum_{m=-\infty}^{m=\infty}\left[\dfrac{\coefficients_{\mathrm{\left(1\right)}m}(\Blochwn)\coefficients_{\mathrm{\left(2\right)}-m}(-\Blochwn)}{\adjointerm_{m}^{+}}+\dfrac{\coefficients_{\mathrm{\left(1\right)-m}}(-\Blochwn)\coefficients_{\mathrm{\left(2\right)}m}(\Blochwn)}{\adjointerm_{m}^{-}}\right]+\ensemble{\coefficients_{\mathrm{\left(1\right)}}},\nonumber \\
\ensemble{\rho(\positionscalar)\scalarG}(\xi) & = & \coeffG\sum_{m=-\infty}^{m=\infty}\left[\dfrac{\rho_{m}(\Blochwn)a_{-m}(-\Blochwn)}{\adjointerm_{m}^{+}}+\dfrac{\rho_{-m}(-\Blochwn)a_{m}(\Blochwn)}{\adjointerm_{m}^{-}}\right],\\
\ensemble{\rho(\positionscalar)\scalarG\rho(\positionscalar^{\prime})}(\xi) & = & \coeffG\sum_{m=-\infty}^{m=\infty}\left[\dfrac{\rho_{m}(\Blochwn)\rho_{-m}(-\Blochwn)}{\adjointerm_{m}^{+}}+\dfrac{\rho_{-m}(-\Blochwn)\rho_{m}(\Blochwn)}{\adjointerm_{m}^{-}}\right],
\end{eqnarray}
where $\coefficients_{\mathrm{\left(1\right)}}$ and $\coefficients_{\mathrm{\left(2\right)}}$
denote two different properties from the set $\left\{ \elascalar,\check{\elascalar},\dfrac{\Piezoelectricscalar}{\Permittivityscalar},\dfrac{\Piezoelectricscalar^{2}}{\Permittivityscalar}\right\} $,
and 
\[
\adjointerm_{m}^{\pm}=i\left(\Blochwn-\dfrac{2\pi m}{l}\pm\xi\right).
\]
Note that

\begin{eqnarray}
\ensemble{\scalarG{}_{,\positionscalar^{\prime}}\rho(\positionscalar^{\prime})}(\xi) & = & \ensemble{\rho(\positionscalar)\scalarG{_{,\positionscalar}}}(-\xi),\\
\ensemble{\coefficients_{\mathrm{\left(2\right)}}(\positionscalar)\scalarG{}_{,\positionscalar^{\prime}}\coefficients_{\mathrm{\mathrm{\left(1\right)}}}(\positionscalar^{\prime})}(\xi) & = & \ensemble{\coefficients_{\mathrm{\left(1\right)}}(\positionscalar)\scalarG{}_{,\positionscalar}\coefficients_{\mathrm{\mathrm{\left(2\right)}}}(\positionscalar^{\prime})}(-\xi),\\
\ensemble{\scalarG\rho(\positionscalar^{\prime})}(\xi) & = & \ensemble{\rho(\positionscalar)\scalarG}(-\xi).
\end{eqnarray}
The process is exemplified using the calculation of $\ensemble{\elascalar(\positionscalar)\,\scalarG{_{,\positionscalar}}}(\xi)$.
Firstly, note that in realization $y=0$ is 
\begin{eqnarray}
\zerosub{\elascalar}(\positionscalar)\scalarG_{0,x}=\left\lbrace \begin{array}{l}
\coeffG\zerosub{\elascalar}(\positionscalar)\pl u_{,x}(\positionscalar)\mn u(\positionscalar^{\prime}),\;\;\;\positionscalar<\positionscalar^{\prime},\\[10pt]
\coeffG\pl u(\positionscalar^{\prime})\zerosub{\elascalar}(\positionscalar)\mn u_{,x}(\positionscalar),\;\;\;\positionscalar^{\prime}<\positionscalar.
\end{array}\right.
\end{eqnarray}

\noindent The translated expression is thus 
\begin{equation}
\elascalar(\positionscalar)\scalarG{_{,\positionscalar}}=\left\lbrace \begin{array}{l}
\coeffG e^{i\Blochwn(\positionscalar-\positionscalar^{\prime})}\sum\limits _{m=-\infty}^{m=\infty}C_{m}(\Blochwn)e^{\frac{-2i\pi m(\positionscalar-y)}{l}}\sum\limits _{n=-\infty}^{n=\infty}a_{n}(-\Blochwn)e^{\frac{-2i\pi n(\positionscalar^{\prime}-y)}{l}},\;\;\;\positionscalar<\positionscalar^{\prime}\\[10pt]
\coeffG e^{i\Blochwn(\positionscalar^{\prime}-\positionscalar)}\sum\limits _{m=-\infty}^{m=\infty}a_{m}(\Blochwn)e^{\frac{-2i\pi m(\positionscalar^{\prime}-y)}{l}}\sum\limits _{n=-\infty}^{n=\infty}C_{n}(-\Blochwn)e^{\frac{-2i\pi n(\positionscalar-y)}{l}}\;\;\;\positionscalar^{\prime}<\positionscalar.
\end{array}\right.\label{eq:translatedCG}
\end{equation}
It follows that 
\begin{eqnarray}
\ensemble{\elascalar(\positionscalar)\,\scalarG{_{,\positionscalar}}}\,(\positionscalar-\positionscalar^{\prime})=\left\lbrace \begin{array}{l}
\coeffG e^{i\Blochwn(\positionscalar-\positionscalar^{\prime})}\sum\limits _{m=-\infty}^{m=\infty}C_{m}(\Blochwn)a_{-m}(-\Blochwn)e^{\frac{2i\pi m(\positionscalar^{\prime}-\positionscalar)}{l}},\;\;\;\positionscalar<\positionscalar^{\prime}\\[10pt]
\coeffG e^{i\Blochwn(\positionscalar^{\prime}-\positionscalar)}\sum\limits _{m=-\infty}^{m=\infty}a_{m}(\Blochwn)C_{-m}(-\Blochwn)e^{\frac{2i\pi m(\positionscalar-\positionscalar^{\prime})}{l}}\;\;\;\positionscalar^{\prime}<\positionscalar,
\end{array}\right.
\end{eqnarray}
and its Fourier transform is

\noindent 
\begin{equation}
\ensemble{\elascalar(\positionscalar)\scalarG{_{,\positionscalar}}}(\xi)=\coeffG\sum_{m=-\infty}^{m=\infty}\left[\dfrac{C_{m}(\Blochwn)a_{-m}(-\Blochwn)}{i\left(\Blochwn-\dfrac{2\pi m}{l}+\xi\right)}+\dfrac{a_{m}(\Blochwn)C_{-m}(-\Blochwn)}{i\left(\Blochwn-\dfrac{2\pi m}{l}-\xi\right)}\right].\label{Example-1}
\end{equation}

\global\long\def\theequation{D.\arabic{equation}}%
 \setcounter{equation}{0}

\section*{\label{Appendix D}Appendix D. Explicit expressions for the effective
properties}

Recall that in the present problem 
\begin{equation}
\stressindex=\left(\elascalar+\dfrac{\Piezoelectricscalar^{2}}{\Permittivityscalar}\right)\left(u_{,x}-\eta\right),\;\;\;\phi{}_{,x}=\dfrac{\Piezoelectricscalar}{\Permittivityscalar}\left(u_{,x}-\eta\right),\;\;\;\Edisplaindex=0,\;\;\;\scalarmomentum=s\rho u,\label{properties-medium}
\end{equation}
and hence 
\begin{equation}
\ensemble{\stressindex}=\ensemble{\left(\elascalar+\dfrac{\Piezoelectricscalar^{2}}{\Permittivityscalar}\right)\left(u_{,x}-\eta\right)},\;\;\;\ensemble{\phi_{,x}}=\ensemble{\dfrac{\Piezoelectricscalar}{\Permittivityscalar}\left(u_{,x}-\eta\right)},\;\;\;\ensemble{\Edisplaindex}=0,\;\;\;\ensemble{\scalarmomentum}=\ensemble{s\rho u}.\label{mean-properties}
\end{equation}
In terms of the effective properties, we we also have that 
\begin{eqnarray}
\begin{aligned}\ensemble{\stressindex} & =\left(\effective{\elascalar}+\dfrac{\tp{\tilde{\Piezoelectricscalar}}\effective{\Piezoelectricscalar}}{\effective{\Permittivityscalar}}\right)\left(\ensemble{u_{,x}}-\eta\right)+\left(\effective{\williscalar}+\tp{\tilde{\Piezoelectricscalar}}\dfrac{\effective{\rgscalar}}{\effective{\Permittivityscalar}}\right)\ensemble{su},\\[3pt]
\ensemble{\phi_{,x}} & =\dfrac{\effective{\Piezoelectricscalar}}{\effective{\Permittivityscalar}}\left(\ensemble{u_{,x}}-\eta\right)+\dfrac{\effective{\rgscalar}}{\effective{\Permittivityscalar}}\ensemble{su},\\[3pt]
\ensemble{\scalarmomentum} & =\left(\adjoint{\effective{\williscalar}}+\adjoint{\effective{\rgscalar}}\dfrac{\effective{\Piezoelectricscalar}}{\effective{\Permittivityscalar}}\right)\left(\ensemble{u_{,x}}-\eta\right)+\left(\effective{\rho}+\dfrac{\effective{\rgscalar}\adjoint{\effective{\rgscalar}}}{\effective{\Permittivityscalar}}\right)\ensemble{su}.
\end{aligned}
\label{mean-properties-Willis}
\end{eqnarray}
By substituting

\begin{eqnarray}
\begin{aligned}u\left(x,y\right)= & \left(G\ensemble G{}^{-1}\ensemble{G_{,x^{\prime}}\check{\scalarelas}}-G{}_{,x^{\prime}}\check{\scalarelas}\right)\left(\ensemble{u_{,x'}}-\eta\right)\\
 & +\left(G\ensemble G{}^{-1}\ensemble{G\rho}-G\rho\right)s^{2}\ensemble u+\ensemble u,
\end{aligned}
\label{u-WillisFormulation-1}
\end{eqnarray}
into Eq.~\eqref{mean-properties} and comparing with Eq.~\eqref{mean-properties-Willis}
we obtain

\begin{eqnarray}
\effective{\elascalar}+\dfrac{\tp{\tilde{\Piezoelectricscalar}}\effective{\Piezoelectricscalar}}{\effective{\Permittivityscalar}} & = & \ensemble{\check{\elascalar}}-\ensemble{\check{\elascalar}(\positionscalar)\scalarG{}_{,\positionscalar\positionscalar^{\prime}}\check{\elascalar}(\positionscalar^{\prime})}+\ensemble{\check{\elascalar}(\positionscalar)\scalarG{}_{,\positionscalar}}\ensemble{\scalarG}^{-1}\ensemble{\scalarG{}_{,\positionscalar^{\prime}}\check{\elascalar}(\positionscalar^{\prime})},\nonumber \\
\effective{\williscalar} & = & -s\ensemble{\elascalar(\positionscalar)\scalarG{}_{,\positionscalar}\rho(x^{\prime})}+s\ensemble{\elascalar(\positionscalar)\scalarG{}_{,\positionscalar}}\ensemble{\scalarG}^{-1}\ensemble{\scalarG\rho(\positionscalar^{\prime})},\nonumber \\
\tp{\tilde{\Piezoelectricscalar}}\dfrac{\effective{\rgscalar}}{\effective{\Permittivityscalar}} & = & -s\ensemble{\dfrac{\Piezoelectricscalar(x)^{2}}{\Permittivityscalar(x)}\scalarG{}_{,\positionscalar}\rho(x^{\prime})}+s\ensemble{\dfrac{\Piezoelectricscalar(x)^{2}}{\Permittivityscalar(x)}\scalarG{}_{,\positionscalar}}\ensemble{\scalarG}^{-1}\ensemble{\scalarG\rho(\positionscalar^{\prime})},\label{eq:Const-operators-1}
\end{eqnarray}
from $\ensemble{\stressindex}$; from $\ensemble{\phi_{,x}}$ we find

\begin{eqnarray}
\dfrac{\effective{\Piezoelectricscalar}}{\effective{\Permittivityscalar}} & = & \ensemble{\dfrac{\Piezoelectricscalar}{\Permittivityscalar}}-\ensemble{\dfrac{\Piezoelectricscalar(x)}{\Permittivityscalar(x)}\scalarG{}_{,\positionscalar\positionscalar^{\prime}}\check{\elascalar}(\positionscalar^{\prime})}+\ensemble{\dfrac{\Piezoelectricscalar(x)}{\Permittivityscalar(x)}\scalarG{}_{,\positionscalar}}\ensemble{\scalarG}^{-1}\ensemble{\scalarG{}_{,\positionscalar^{\prime}}\check{\elascalar}(x^{\prime})},\nonumber \\
\dfrac{\effective{\rgscalar}}{\effective{\Permittivityscalar}} & = & -s\ensemble{\dfrac{\Piezoelectricscalar(x)}{\Permittivityscalar(x)}\scalarG{}_{,\positionscalar}\rho(x^{\prime})}+s\ensemble{\dfrac{\Piezoelectricscalar(x)}{\Permittivityscalar(x)}\scalarG{}_{,\positionscalar}}\ensemble{\scalarG}^{-1}\ensemble{\scalarG\rho(\positionscalar^{\prime})},\label{eq:Const-operators-2}
\end{eqnarray}
and finally from $\ensemble{\scalarmomentum}$ 
\begin{eqnarray}
\adjoint{\effective{\williscalar}} & = & -s\ensemble{\rho(\positionscalar)\scalarG{}_{,\positionscalar^{\prime}}\elascalar(\positionscalar^{\prime})}+s\ensemble{\rho(\positionscalar)\scalarG}\ensemble{\scalarG}^{-1}\ensemble{\scalarG{}_{,\positionscalar^{\prime}}\elascalar(\positionscalar^{\prime})},\nonumber \\
\adjoint{\effective{\rgscalar}}\dfrac{\effective{\Piezoelectricscalar}}{\effective{\Permittivityscalar}} & = & -s\ensemble{\rho(\positionscalar)\scalarG{}_{,\positionscalar^{\prime}}\dfrac{\Piezoelectricscalar(\positionscalar^{\prime})^{2}}{\Permittivityscalar(\positionscalar^{\prime})}}+s\ensemble{\rho(\positionscalar)\scalarG}\ensemble{\scalarG}^{-1}\ensemble{\scalarG{}_{,\positionscalar^{\prime}}\dfrac{\Piezoelectricscalar(\positionscalar^{\prime})^{2}}{\Permittivityscalar(\positionscalar^{\prime})}},\nonumber \\
\effective{\rho}+\dfrac{\effective{\rgscalar}\adjoint{\effective{\rgscalar}}}{\effective{\Permittivityscalar}} & = & \ensemble{\rho}-s^{2}\ensemble{\rho(\positionscalar)\scalarG\rho(\positionscalar^{\prime})}+s^{2}\ensemble{\rho(\positionscalar)\scalarG}\ensemble{\scalarG}^{-1}\ensemble{\scalarG\rho(\positionscalar^{\prime})}.\label{eq:Const-operators-3}
\end{eqnarray}
As we pointed out in the body of the paper 
\begin{align}
\effective{\williscalar}\left(\xi\right) & =\adjoint{\effective{\williscalar}}\left(-\xi\right)=-\mathrm{conj}\,\adjoint{\effective{\williscalar}}\left(\xi\right).\label{eq:conj S}
\end{align}
Also note that $\dfrac{\tp{\tilde{\Piezoelectricscalar}}\effective{\rgscalar}}{\effective{\Permittivityscalar}}\left(\xi\right)=\dfrac{\effective{\Piezoelectricscalar}\adjoint{\effective{\rgscalar}}}{\effective{\Permittivityscalar}}\left(-\xi\right)$.
As mentioned, the drawback for using a single Green function in the
absence of charge is one degree of freedom in calculating $\effective{\lmat}$,
which we eliminate by enforcing $\tp{\tilde{\Piezoelectricscalar}}=\effective{\Piezoelectricscalar}$.
The latter condition, together with Eqs.~\eqref{eq:Const-operators-1}-\eqref{eq:Const-operators-3}
deliver the effective properties $\effective{\elascalar},\effective{\Permittivityscalar},\effective{\rho},\effective{\Piezoelectricscalar}$
and $\effective{\rgscalar},\adjoint{\effective{\rgscalar}}$. These
satisfy 
\begin{align}
\tilde{c}\left(\xi\right)= & \mathrm{conj}\,\tilde{c}\left(-\xi\right),\quad\tilde{c}=\effective{\elascalar},\effective{\Permittivityscalar},\effective{\rho},\effective{\Piezoelectricscalar},\label{eq:properties conj}\\
\effective{\rgscalar}\left(\xi\right) & =\adjoint{\effective{\rgscalar}}\left(-\xi\right)=-\mathrm{conj}\,\adjoint{\effective{\rgscalar}}\left(\xi\right),\label{eq:W conj}
\end{align}
where relation between $\effective{\rgscalar}$ and $\adjoint{\effective{\rgscalar}}$
is similar to the relation between $\effective{\williscalar}$ and
$\adjoint{\effective{\williscalar}}$.

\setcounter{equation}{0}


\bibliographystyle{unsrt}
\bibliography{bibtexfiletot}
\end{document}